\documentclass[11pt]{article}
\usepackage[margin=1in]{geometry}
\usepackage{amsmath,amssymb,amsthm,mathtools}
\usepackage{graphicx}
\usepackage{tikz}
\usetikzlibrary{positioning,arrows.meta,fit,backgrounds}
\usepackage{booktabs}
\usepackage{array}
\usepackage{caption}
\usepackage{float}
\usepackage{enumitem}
\usepackage{bm}
\usepackage{xcolor}
\usepackage[colorlinks=true,linkcolor=blue,citecolor=blue,urlcolor=blue]{hyperref}
\usepackage{natbib}

\newcommand{\E}{\mathbb{E}}
\newcommand{\PR}{\mathbb{P}}
\newcommand{\R}{\mathcal{R}}
\newcommand{\AAL}{\mathrm{AAL}}
\newcommand{\VaR}{\mathrm{VaR}}
\newcommand{\TVaR}{\mathrm{TVaR}}
\newcommand{\BEP}{\mathrm{BEP}}
\newcommand{\BAEP}{\mathrm{BAEP}}
\newcommand{\BOEP}{\mathrm{BOEP}}
\newcommand{\OEP}{\mathrm{OEP}}
\newcommand{\AEP}{\mathrm{AEP}}
\newcommand{\EP}{\mathrm{EP}}
\newcommand{\Cat}{\mathrm{Cat}}
\newcommand{\Par}{\mathrm{Par}}
\newcommand{\Corr}{\mathrm{Corr}}
\newcommand{\Cov}{\mathrm{Cov}}
\newcommand{\Var}{\mathrm{Var}}

\theoremstyle{plain}

\newtheorem{remark}{Remark}

\title{NatPar: Natural Parametric Modeling\thanks{\copyright\ Hirbod Assa}}
\author{Hirbod Assa\thanks{Model Library and UCD, email: assa.hirbod@gmail.com}}
\date{\today}

\begin{document}
\maketitle

\begin{abstract}
We develop \emph{natural parametric} (NatPar) insurance as the natural next step from natural-catastrophe
(NatCat) modelling: the same hazard--exposure--vulnerability--finance machinery, with a parametric
index made contractual in place of indemnity loss adjustment. Our aim is practical---to establish a
\emph{standard approach}, inspired by how NatCat models and the catastrophe-insurance industry
already operate, rather than to propose another optimal-contract criterion. This vantage point
delivers two payoffs. First, it fixes \emph{how the reporting is formulated}: NatPar contracts are
reported in the native NatCat language---annual average loss, EP/AEP/OEP curves, return-period
levels, one-year tail metrics---and complemented with two-sided basis-exceedance diagnostics
($\BEP^\pm$ and portfolio basis AEP/OEP) that we elevate to the same central status the EP curve
holds for losses: $\BEP^\pm$ is the canonical distributional view of basis risk, not a
supplementary number. We package these into a minimal reporting template that makes
designs comparable, auditable, and usable by underwriting, capital, and supervisory functions rather
than living only in abstract contract-design space. Second, the same standard lets us see \emph{how
the tail is reallocated} between insuree and insurer. As a pricing tool we pose trigger pricing as a
single loss-minimisation over the two-sided basis, shaping the premium through both a weight and a
\emph{curvature} on shortfall versus overpayment; AAL-neutral (``fair'') pricing is the
symmetric-linear corner of this problem, a special case rather than a separate paradigm.

A frost case study with three simulated regions sharing one identical published damage function illustrates the
framework and yields its central result, which is about \emph{time rather than average}. Holding a
contract AAL-neutral, the bounded payout cannot follow the unbounded exposure-severity tail, so
equalising the mean forces over- and under-payment to separate across return periods: the insuree
gains at short and moderate horizons while the insurer sheds the deep tail past a crossover of
several decades. This relocation sharpens as exposure growth and climate change fatten the loss tail, but
reverses once hazards exhibit \emph{tail dependence}---when regions reach their extremes jointly,
bounded payouts stack and the insurer reabsorbs the deep tail. The incidence of catastrophe risk
under NatPar is thus the resultant of two opposing climate forces, neither visible to a
correlation-based portfolio model, giving a precise account of why bounded parametric structures
have grown in popularity and where their governance attention belongs.
\end{abstract}

\tableofcontents

\section{Introduction}\label{sec:intro}
Over the last decade, climate pressure alongside rapid economic growth has increased both
exposure and hazard intensity across many regions. In several markets, climate-exposed
lines have moved toward outcomes increasingly described---informally or explicitly---as
uninsurable: coverage is withdrawn, terms are tightened, or prices rise to levels that are
politically and socially untenable. At the same time, protection gaps for climate perils
remain large and, in some settings, continue to widen.

From the perspective of insurers and reinsurers, the binding challenge is often not only
higher mean losses, but reduced tractability of the loss distribution, especially in the tail.
Hazard distributions drift, exposures evolve quickly, and extremes may cluster in space
and time; these dynamics inflate uncertainty precisely where solvency frameworks and risk
appetite concentrate. Long-tailed indemnity liabilities become harder to reserve, harder
to explain, and harder to support under one-year capital constraints and target returns.
In this sense, a line drifts toward uninsurability when markets cease to clear because one
or more constraints becomes binding: affordability (premiums exceed willingness or ability
to pay), capital (tail requirements render the line uneconomic), and/or model uncertainty
(pricing and reserving assumptions are no longer defensible to management, regulators,
or capacity providers). These constraints reinforce each other: uncertainty in exposure
and vulnerability increases capital charges and risk loadings, pushing premiums higher and
accelerating capacity withdrawal.

Against this background, parametric structures have expanded not merely as customer-facing innovations, but as a supply-side reconfiguration of NatCat practice. Under climate-driven uninsurability and capital pressure, NatCat teams increasingly redeploy the hazard--exposure--vulnerability--finance stack to produce index-linked liabilities that are auditable,
bounded by contract, and more capital-tractable. We refer to this pipeline as Natural
Parametric (NatPar) modelling. Conceptually, NatPar preserves the NatCat architecture
but changes what becomes contractual: hazard modelling defines measurable indices and
triggers; exposure and vulnerability are re-tasked into a basis-risk engine that quantifies the
mismatch between realised indemnity loss and parametric payout; and the finance block
is adapted to short-tailed liabilities with minimal development risk. NatPar therefore exchanges a portion of exposure-driven indemnity variability for a transparent, index-verified
payout, reducing reserve uncertainty and improving capital tractability at the cost of explicit basis risk.

\paragraph{Design-first versus pipeline-first.} It is useful to contrast our objective---to set standards for NatPar reporting and regulation--- with the dominant framing in much of
the academic literature and industrial development.
\begin{itemize}[leftmargin=1.4em]
\item \textbf{Academic view: design-first.} A large literature treats parametric insurance as a
contract-design problem: specify an index (or index vector) $X$ and choose a payout
function $I(X)$ to optimise a welfare or risk objective under a premium constraint.
This perspective is mathematically flexible and emphasises basis risk as the mismatch
between realised loss and index-based payout, making it well suited for studying
optimality, robustness, behavioural frictions, and uptake.
\item \textbf{Industry view: pipeline-first.} In practice, however, many parametric products are
developed inside the NatCat ecosystem. Underwriting, reinsurance/ILS structuring,
accumulation control, and solvency reporting rely on a modular workflow (hazard,
exposure, vulnerability, financial terms) and a standard set of portfolio outputs (AAL
and EP/AEP/OEP curves, return-period levels, and one-year tail metrics). From
this viewpoint, a parametric contract is not an arbitrary $I(X)$: it must be verifiable,
auditable, and comparable within the same reporting language used for indemnity
portfolios and capital.
\end{itemize}
This creates a practical gap between design-first theory and pipeline-first implementation---a gap that becomes acute when products must be justified to boards, supervisors, and
capacity providers.

This paper takes the view that NatPar is the \emph{natural next step from NatCat}: not a new
object grafted onto catastrophe modelling, but the same hazard--exposure--vulnerability--finance
machinery with a different question asked of it---what should become contractual. Our aim is
correspondingly practical. Rather than proposing yet another optimal-contract criterion, we set out
to \textbf{establish a standard approach}---one deliberately inspired by how NatCat models and the
catastrophe-insurance industry already work---so that parametric contracts can be built, reported,
and governed with the tools practitioners already use. The contribution is therefore methodological
and infrastructural rather than a single theorem: a way of working that is comparable to indemnity
practice, auditable by supervisors, and usable by underwriting and capital functions. Adopting this
vantage point delivers two concrete payoffs, around which the paper is organised: it fixes
\emph{how the reporting is formulated}, and it lets us see \emph{how the tail is reallocated}
between the parties.

Both payoffs rest on a single methodological departure. Almost the entire NatCat apparatus carries
over untouched---the hazard, exposure, and vulnerability blocks, the AAL and EP/AEP/OEP outputs, the
capital metrics. The one thing that is genuinely new when an index becomes contractual is the
mismatch between the indemnity loss and the parametric payout, the \emph{basis risk}, and our
central claim is about how to look at it: as a \emph{distribution}, reported with the same
exceedance-and-return-period machinery already used for losses, rather than collapsed into a single
correlation or hedge-effectiveness number. This distributional reading of basis risk is what makes
the reporting standard auditable and what makes the tail-reallocation result visible; the two
payoffs below are its two consequences.

The first payoff is \textbf{how the reporting is formulated}: a standard built on the basis-risk
distribution. Because NatPar lives inside the NatCat stack, its parametric contracts are constructed
and reported with the conventional portfolio outputs that the industry already produces---AAL,
EP/AEP/OEP curves, return-period levels, and one-year tail metrics---so they are directly comparable
to indemnity programmes rather than living in a separate analytical world. The one genuinely new
object is basis risk, and the practical choice we make is to report it not as a single summary
number but as a \emph{distribution}, through two-sided basis-exceedance curves ($\BEP^\pm$) for
shortfall and overpayment, with their portfolio (BAEP/BOEP) and return-period counterparts. The
emphasis matters: we give $\BEP^\pm$ the same central status that the EP curve has on the loss side.
It is the canonical view of basis risk---the object the report is built around and the one against
which contract shapes are compared---not a supplementary tail diagnostic bolted on beside the loss
EP curve. Wherever the existing workflow reaches for an EP curve, the NatPar workflow reaches for
the matching $\BEP^\pm$. We compute these under an AAL-neutral calibration
so that the shortfall/overpayment structure reflects payout shape rather than a difference in mean
loss, and we package the whole into a minimal reporting template---a concrete, repeatable checklist
that makes NatPar designs comparable, auditable, and amenable to supervision.

The second payoff is that the same standard lets us see \textbf{how the tail is reallocated}.
Holding contracts AAL-neutral isolates the effect of payout shape on the tail: contracts with the
same mean loss can have materially different tails, so AAL-neutrality does not imply
capital-neutrality under VaR-style solvency measures. More strikingly, the bounded parametric payout
cannot follow the unbounded exposure-severity tail of the indemnity loss, so equalising the mean
\emph{relocates} mismatch across return periods rather than removing it. This is a concrete,
reportable statement about the \emph{allocation of risk between insuree and insurer} over time---the
insuree tends to gain at short horizons while the insurer sheds the deep tail at long ones---and,
because it is expressed in the same return-period language as the rest of the report, it is exactly
the kind of thing a board or regulator can read off directly. We show how diversification and,
crucially, tail dependence among hazards move that allocation.

Underpinning both payoffs is a pricing tool we develop in passing but use throughout:
\textbf{trigger pricing posed as loss-minimisation}. The premium is the solution of minimising the
expected disutility of the two-sided basis, shaped by both a weight and a \emph{curvature} on
shortfall versus overpayment (more generally, any convex loss). Asymmetric curvature relocates the
premium along the shortfall/overpayment trade-off; the symmetric-linear corner of this problem
coincides with AAL-neutral (``fair'') pricing, which we therefore treat as a special case rather
than a separate paradigm. We illustrate all of this in a single case study with three simulated regions sharing one identical damage function, applying a consistent reporting template throughout.

The remainder of the paper is organised as follows. Section~\ref{sec:lit} reviews the contract-design
and multi-hazard index-insurance literatures and contrasts them with the pipeline and reporting language used in catastrophe practice. Section~\ref{sec:natcat} summarises the NatCat hazard--exposure--vulnerability--finance stack and the standard portfolio outputs used for underwriting and capital. Section~\ref{sec:natpar} then formalises the NatPar mapping and introduces the minimal reporting
template. Section~\ref{sec:pricing} poses trigger pricing as a loss-minimisation over the basis.
Sections~\ref{sec:blocks}, \ref{sec:frost} and \ref{sec:interp}
implement the framework in a frost case study with three simulated regions sharing one identical damage function:
Section~\ref{sec:blocks} develops the analytic
representations, Section~\ref{sec:frost} reports the empirical
outputs for both individual risks and a three-region portfolio, and
Section~\ref{sec:interp} interprets the results. Section~\ref{sec:ext} discusses extensions,
regulatory implications, and practical limitations, and Section~\ref{sec:summary} concludes.

\section{Literature review}\label{sec:lit}
Much of the academic literature treats parametric insurance as a generic contract-design
problem, whereas the industry has largely evolved parametric programmes as pipeline-compatible overlays on NatCat machinery.

A standard academic entry point models insurance design as a choice of a payout schedule subject to pricing constraints, often derived from expected-utility considerations and
classic optimal insurance results (e.g., generalized deductibles) \citep{arrow1974,raviv1979}. Index (parametric)
insurance adopts a related contract-design perspective, but replaces loss adjustment by a
payout determined by an observable index (or index vector) $X$, typically written as $I(X)$.
Modern treatments formalize the choice of $I(\cdot)$ under premium constraints and study welfare, risk, and implementability of index contracts \citep{zhang2019}. Recent surveys situate these design principles within the broader agenda of quantitative agricultural risk management \citep{assa2025intro}, and \citet{zhu2025index} give a comprehensive review of index-insurance design---its actuarial framework, empirical findings, and emerging AI and blockchain tooling---in which basis risk is again the central design constraint.

Across this literature, basis risk---the mismatch between realized indemnity-style loss
and the index-based payout---is consistently identified as the central friction that can dominate performance and suppress demand. Two influential directions are (i) multi-scale
indices to reduce mismatch across spatial/temporal aggregation and (ii) empirical measurement of index quality and basis-risk outcomes in the field \citep{elabed2013,jensen2016}. Complementary work
emphasises implementation and uptake constraints, including regulatory challenges and
behavioural frictions such as complexity aversion, which can reduce uptake even when
expected values are favorable \citep{sonsino2001,sonsino2002}.

A conceptually close strand to our motivation is the review by \citet{benso2023} on
weather index insurance for multi-hazard resilience and food security. Beyond synthesizing
the literature, they propose a practical three-module framework for index insurance design:
hazard identification, vulnerability assessment, and financial methods and risk pricing. This
decomposition is a useful bridge between a broad (and sometimes fragmented) academic
literature and the practical steps required to build indices and payout rules.

A key message in the multi-hazard setting is that hazard representation and combination are modelling choices, not merely data choices. \citet{benso2023} emphasise that multi-hazard products should not default to independence assumptions and highlight interaction
taxonomies (independent, synergistic, cascading) that can materially affect loss modelling
and premium adequacy \citep{benso2023,gill2014}. Related work reviews quantitative methodologies for multi-hazard interrelationships and provides a structured view of how hazard interactions can be
represented and tested \citep{tilloy2019}. In climate-risk contexts, compound-event research provides
additional conceptual and statistical tools for thinking about multivariate extremes and
interacting drivers \citep{zscheischler2020}.

In industry practice, catastrophe risk is commonly represented via a modular supply
chain: hazard (event sets/footprints), exposure (assets at risk and where), vulnerability
(damage/loss functions), and financial terms (deductibles, limits, aggregates and layers).
This architecture is central to how portfolios are priced (e.g., AAL) and how tails are
communicated (e.g., EP/AEP/OEP curves and return-period losses), and it is described
in practitioner-oriented treatments of catastrophe risk modelling \citep{mitchellwallace2017}.
On the policy side, index-based schemes are increasingly embedded in public risk-management
frameworks; \citet{santeramo2025eu} analyse weather and yield index instruments in EU agriculture
and the Agri-CAT fund, illustrating how index triggers and mutual-fund structures combine in
practice.

Recent climate physical risk assessment (PRA) and climate change risk assessment
(CCRA) work builds naturally on the same decomposition, but adds scenario conditioning, spatial-temporal resolution constraints, and climate-model uncertainty. In particular,
portfolio-scale CCRA requires separating changes in hazard from changes in exposure, because their interaction can magnify or attenuate losses \citep{bourget2024}. Integrating climate scenarios
into NatCat-style workflows typically requires explicit choices around downscaling and bias
correction, which can materially influence hazard projections and therefore pricing and capital metrics \citep{chen2021,maraun2018,martynov2013,martel2020}. At the same time, climate disclosure and supervisory stress
testing have accelerated demand for portfolio-level methodologies that are transparent and
auditable \citep{fsb2017,boe2019,osfi2023}.

The contract-design literature explains parametric insurance in the abstract language
of $I(X)$ and basis risk, and multi-hazard reviews provide useful design taxonomies \citep{zhang2019,benso2023}.
However, these literatures do not typically track how parametric products have evolved
operationally inside the insurance ecosystem: namely, as contractual overlays on the
catastrophe-modelling pipeline that already governs underwriting, portfolio aggregation,
reinsurance/ILS structuring, and solvency reporting.

This paper therefore adopts a pipeline-first notion of parametric insurance. We define
NatPar programmes as contracts whose triggers, payouts, and validation procedures are
explicitly designed to be compatible with the NatCat supply chain and its portfolio outputs.
In this view, basis risk is not merely a single summary statistic but the measurable interface
created when (exposure $\times$ vulnerability) is replaced by a low-dimensional, index-verifiable
hazard proxy. This is also where model governance lives: NatPar design should be readable
in the same reporting language as NatCat (AAL, EP/AEP/OEP, tail metrics), with explicit
basis distributions and basis exceedance curves as diagnostics.

\section{NatCat machinery in practitioner terms}\label{sec:natcat}
This section summarises the catastrophe-modelling workflow that underpins NatCat underwriting and portfolio reporting. We focus on the modelling blocks and the portfolio outputs
that are used in practice, because NatPar is defined in this paper as a pipeline-compatible
overlay on the same machinery.

\subsection{The four modeling blocks: hazard, exposure, vulnerability, financial terms}
A standard NatCat model can be represented as a modular pipeline:
\[
\text{Hazard} \;\to\; \text{Exposure} \;\to\; \text{Vulnerability} \;\to\; \text{Financial terms}.
\]
The purpose of this decomposition is operational: each block can be updated, validated,
and governed with its own data sources and model risk controls, while still producing
consistent portfolio outputs.

\paragraph{Hazard.} The hazard block specifies an event set (or stochastic process) for the peril,
including frequency, intensity, and spatial footprints where relevant. In climate settings,
the hazard block may also be conditioned on scenarios or nonstationary assumptions.

\paragraph{Exposure.} The exposure block specifies what is at risk and where: locations, sums
insured/values, relevant asset attributes, and aggregation rules. Exposure is often the
most rapidly changing component in practice, and it is also a major driver of accumulation
risk.

\paragraph{Vulnerability.} The vulnerability block maps hazard intensity (and modifiers) to damage ratios or loss distributions. It captures construction/asset sensitivity, secondary uncertainty, and model error.

\paragraph{Financial terms.} Financial terms transform ground-up loss into contractual loss: deductibles, limits, layers, aggregates, reinstatements, and other features that define the
insurer or reinsurer liability.

\subsection{Portfolio loss}
Let $Y_r$ denote the contractual loss after applying financial terms for the set of covered risk
$r \in \R$. Here $\R$ is a set of risks, that can be for example a deterministic set of regions (as
we use in this paper), or even a random set of events (such as $r \in \R \equiv \{1 \le r \le N_t\}$ for a
Poisson process $N_t$).

Portfolio-level reporting commonly distinguishes:
\[
\text{Occurrence loss: } M := \max_r Y_r, \qquad \text{Aggregate loss: } S := \sum_r Y_r.
\]
These are the random quantities that generate standard NatCat outputs such as OEP and
AEP curves. NatCat portfolio reporting distinguishes between exceedance probabilities for annual
occurrence loss (OEP) and annual aggregate loss (AEP). Using the annual variables the
one-year curves are defined as
\[
\OEP(x) := \PR(M > x), \qquad \AEP(x) := \PR(S > x), \qquad x \ge 0.
\]
In general, these curves differ because a year can contain multiple loss-causing events:
OEP captures the tail of the single largest event, while AEP captures the tail of the sum
of events. They coincide only in special cases (e.g.\ when there is at most one loss-causing
event per year in the model, or when financial terms collapse event losses into a single
outcome).

\paragraph{Return periods.} A common presentation uses return periods. For any liability $X$, define
the $T$-year return level
\[
x_T(X) := \inf\{x : \PR(X > x) \le 1/T\}.
\]
Return-period curves $T \mapsto x_T(\cdot)$ provide an interpretable mapping from tail probability
to severity. For a return period $T$ (years), the corresponding OEP return-period loss level
$x_T^{\OEP}$ satisfies $\OEP(x_T^{\OEP}) = 1/T$, and similarly for AEP.

\section{NatPar framework}\label{sec:natpar}
\begin{figure}[ht]
\centering
\definecolor{npblue}{HTML}{1F4E79}
\definecolor{npaccent}{HTML}{C0392B}
\begin{tikzpicture}[
   font=\footnotesize,
   box/.style={rectangle, rounded corners, draw=npblue, line width=0.6pt,
               minimum width=2.4cm, minimum height=1.0cm, align=center,
               fill=npblue!8},
   pbox/.style={rectangle, rounded corners, draw=npaccent, line width=0.6pt,
               minimum width=2.4cm, minimum height=1.0cm, align=center,
               fill=npaccent!8},
   flow/.style={-{Stealth[length=2mm]}, line width=0.7pt, draw=npblue},
   pflow/.style={-{Stealth[length=2mm]}, line width=0.7pt, draw=npaccent},
   map/.style={-{Stealth[length=1.6mm]}, line width=0.5pt, draw=gray, dashed},
]
% NatCat row
\node[box] (h)  {Hazard\\\textit{event/index} $T$};
\node[box, right=0.8cm of h] (e) {Exposure\\\textit{value} $A$};
\node[box, right=0.8cm of e] (v) {Vulnerability\\\textit{damage} $D(T)$};
\node[box, right=0.8cm of v] (f) {Financial\\\textit{layer} $Y=(L-l)^+$};
\draw[flow] (h)--(e); \draw[flow] (e)--(v); \draw[flow] (v)--(f);
\node[left=0.25cm of h, text=npblue] {\textbf{NatCat}};
\node[above=0.12cm of e, xshift=1.2cm, text=npblue!80] {$L=A\,D(T)$ (ground-up)};
% NatPar row
\node[pbox, below=2.2cm of h] (ph) {Index \& trigger\\\textit{observable} $T$};
\node[pbox, right=0.8cm of ph] (pb) {Basis-risk engine\\$B=Y-P$};
\node[pbox, right=0.8cm of pb] (pp) {Payout rule\\$P=I(T)$ (bounded)};
\node[pbox, right=0.8cm of pp] (pf) {Finance\\\textit{short-tailed}};
\draw[pflow] (ph)--(pb); \draw[pflow] (pb)--(pp); \draw[pflow] (pp)--(pf);
\node[left=0.25cm of ph, text=npaccent] {\textbf{NatPar}};
% mappings
\draw[map] (h) -- (ph) node[midway, right, text=gray] {\,index logic};
\draw[map] (e) to[out=-90,in=70] node[pos=0.6, right, text=gray] {\,exposure} (pb);
\draw[map] (v) to[out=-110,in=110] node[pos=0.5, left, text=gray] {vulnerability\,} (pb);
\draw[map] (f) -- (pf) node[midway, right, text=gray] {\,no dev.\ tail};
\end{tikzpicture}
\caption{From NatCat to NatPar: the same four-block pipeline, re-tasked. The hazard becomes
the contractual index; the exposure and vulnerability blocks are re-tasked into a basis-risk
engine that measures the mismatch $B=Y-P$ between indemnity loss and parametric payout; and
the finance block becomes short-tailed with limited development risk. The architecture is
unchanged --- what changes is which object becomes contractual.}
\label{fig:natcat2natpar}
\end{figure}
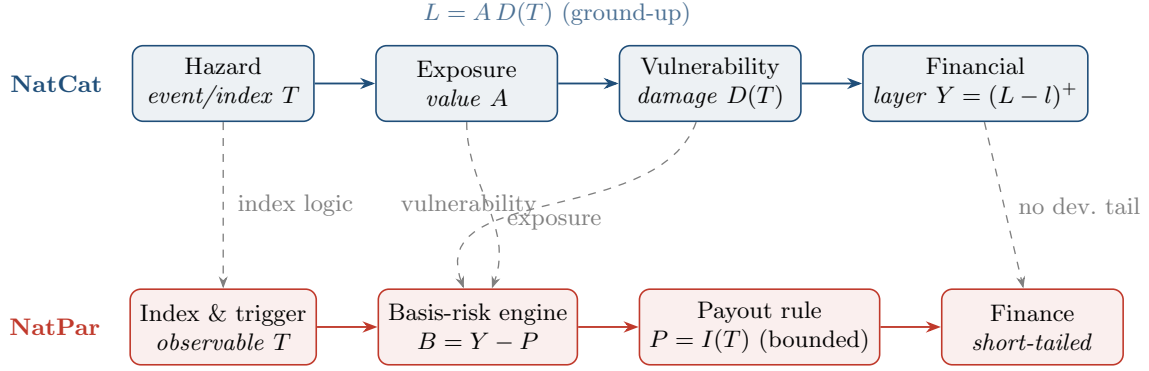

NatPar programmes are defined here as parametric structures anchored in the catastrophe-modelling workflow and evaluated in the same portfolio language as NatCat (AAL, EP/AEP/OEP
curves, return periods, tail metrics). The central additional object is basis risk, which we
treat as an auditable interface rather than an afterthought. Figure~\ref{fig:natcat2natpar}
summarises the mapping: NatPar retains the four-block NatCat pipeline but re-tasks what each
block contributes contractually.

\subsection{NatPar payout classes and their implications for tails and basis}
A practical advantage of a pipeline-first view is that NatPar designs can be organised into
a small number of payout classes, each with predictable implications for tail metrics and
basis risk. This section provides a compact taxonomy used later to interpret the case study.

\paragraph{(i) Continuous bounded payouts (graded contracts).} Contracts of the form $P =
I(X)$ with $I$ continuous and bounded (e.g.\ $P = q(D(X) - d)^+$ with $D \in [0,1]$) mechanically cap the insurer's liability and often reduce high-quantile tail metrics relative to
exposure-scaled indemnity losses, even under AAL-neutral calibration. Their basis risk
tends to be diffuse: frequent moderate mismatch can occur because payouts do not scale
with realised exposure, but severe basis shortfalls can be controlled by the cap and by
shaping $I(\cdot)$.

\paragraph{(ii) Piecewise-linear / tiered payouts.} Piecewise-linear or tiered payouts provide
additional flexibility while remaining transparent and auditable. They can be used to approximate layer-like behaviour (e.g.\ increasing payout rate as hazard severity increases)
and to enforce constraints such as $\VaR_{0.995}(P) \le c$ while still targeting basis shortfall objectives. In a governance setting, the breakpoints of $I(\cdot)$ become explicit design parameters
that can be stress-tested and versioned.

\paragraph{(iii) Trigger-driven events.} A trigger-driven event (or simply binary) $P = q\mathbf{1}\{X\in\text{event}\}$,
for a particular event set and index $X$ for a payout $q$, is maximally simple and highly
verifiable, and they often align with operational objectives such as rapid cashflow upon a
threshold event. However, their loss distribution is discrete, which can have unintuitive
implications: for a trigger probability $1 - \alpha$ (i.e., $\PR\{X \in \text{event}\} = 1 - \alpha$), tail metrics can
``jump'' at confidence levels above $\alpha$ (e.g.\ $\VaR_{0.995}$ may equal $q$ when $\alpha = 0.99$). As a result,
AAL-neutrality does not guarantee capital advantages, and these type of binary designs
can concentrate basis mismatch into rare but large shortfalls or overpayments depending
on how $q$ is calibrated.

\paragraph{(iv) Layer-like parametric designs.} Between continuous and binary extremes lie parametric designs that emulate insurance layers (e.g.\ step functions with multiple thresholds,
or capped linear segments). These are often the most directly comparable to indemnity
layer structures and can be tuned to match both AAL and a tail constraint, while keeping
the trigger logic auditable.

\subsection{NatCat to NatPar through AAL-neutral calibration}
Let $Y$ denote the NatCat contractual loss (for a policy, region, or portfolio segment), and
let $T$ denote an index (typically a hazard proxy or near-hazard proxy) that is observable
and verifiable. A NatPar design specifies a payout rule $P = I(T)$ together with contract
terms such as caps, layers, and trigger thresholds. A natural baseline calibration target is AAL-neutrality:
\[
\E[P] = \E[Y].
\]
This constraint ensures that differences in EP curves and tail metrics primarily reflect distributional shape rather than mean loss. In practice, this can be implemented at different
granularities (policy-level, peril-region segment, or portfolio bucket), and it can be supplemented by loadings for expenses and profit. The key point is that AAL-neutrality is a
comparability device, not a full design criterion.

\begin{remark}[AAL-neutrality relocates tail risk across time, not on average]\label{rem:relocate}
Imposing $\E[P]=\E[Y]$ does not make insurer and insuree indifferent; it only equalises the
\emph{average} transfer. Because $P$ is bounded while $Y$ inherits the unbounded
exposure-severity tail of $A\,D$, holding the mean fixed forces the over- and under-payment
mass to redistribute \emph{across return periods}. The bounded over-payment side concentrates
in frequent, mild years, while the unspanned shortfall side migrates into rare, severe years.
AAL-neutrality is therefore best read not as a fairness statement but as the device that makes
this temporal relocation visible: with the mean held constant, the basis return-period curves of
Section~\ref{sec:frost} isolate \emph{when} each party is ahead. We return to the economic
reading of this relocation, and its link to climate-driven growth in catastrophe risk, in
Section~\ref{sec:whogains} after the numerical results.
\end{remark}

\subsection{Basis risk as an auditable interface}
\emph{This is the paper's major departure.} Everything else---the hazard, exposure, and
vulnerability blocks, the AAL and EP/AEP/OEP outputs, the one-year capital metrics---is inherited
essentially unchanged from NatCat practice. What genuinely changes when an index becomes
contractual is that a new object appears, the mismatch between the indemnity loss and the
parametric payout, and the central methodological claim of this paper is about \emph{how that
object should be looked at}: not as a single summary number (a correlation, an average shortfall, a
hedge-effectiveness ratio), but as a \emph{distribution} in its own right, reported with the same
exceedance-and-return-period machinery already used for losses. Treating basis risk distributionally
is what turns it from an afterthought into a governable, auditable interface, and it is what makes
the downstream adjustments to reporting, underwriting, and portfolio management (developed below and
in Section~\ref{sec:whogains}) possible.

The fundamental trade-off is that $P$ is transparent and bounded but cannot replicate $Y$
pointwise; the resulting mismatch is basis risk. Therefore, define basis risk as
\[
B := Y - P.
\]

\subsubsection{Individual risk basis}
Two useful classes of metric to view a basis risk are:
\begin{itemize}[leftmargin=1.4em]
\item \textbf{Point measures.} Measures that represent the basis risk through a
single scalar such as $\PR(B > 0)$, $\E[B^+]$, $\E[(-B)^+]$, $\Corr(Y, P)$, etc.
\begin{itemize}
\item $\PR(B > 0)$ distinguishes whether the design tends to miss loss states (high shortfall frequency) or to pay in non-loss states (high overpayment frequency).
\item The severities $\E[B^+]$, $\E[(-B)^+]$ are the relevant averages conditional on sign:
they quantify the expected size of liquidity missing when loss occurs (shortfall)
and the expected size of liquidity excess when it is not needed (overpayment).
\item The indemnity-parametric correlation $\Corr(Y, P)$ indicates how the statistical
mismatch between the real payoff and the parametric product can be measured.
\end{itemize}
\item \textbf{The central object: the basis exceedance distribution.} Point summaries are useful
shorthand, but each collapses the mismatch to a single scalar and can hide whether mismatch is
driven by frequent small shortfalls or rare severe ones. The central view of basis risk in this
framework is therefore not a scalar but a \emph{distribution}, reported through its exceedance
curves exactly as losses are reported through the EP curve:
\[
\BEP^+(x) := \PR(B > x), \qquad \BEP^-(x) := \PR(-B > x).
\]
We give these the status of a primary reporting object. \emph{The basis exceedance curve
$\BEP^\pm$ is to basis risk what the EP curve is to loss: the canonical distributional view, not a
supplementary diagnostic.} Just as no catastrophe report would summarise a loss distribution by its
mean alone and omit the EP curve, a NatPar report should not summarise basis risk by a correlation
or an average shortfall and omit $\BEP^\pm$. Every place the workflow currently reaches for an EP
curve, it should now reach for the matching $\BEP^\pm$; the point measures above are read off the
curve, not the other way round.
\end{itemize}

The two views are genuine counterparts. The loss EP curve $\PR(Y>x)$ reports the distribution of
what the buyer must pay; the basis curves $\BEP^\pm$ report the distribution of what the contract
fails to transfer, signed by direction. The shortfall curve $\BEP^+(x) = \PR(B > x)$ answers: how
often does the indemnity benchmark exceed the parametric payout by more than $x$? The overpayment
curve $\BEP^-(x) = \PR(-B > x)$ answers: how often does the parametric payout exceed the indemnity
benchmark by more than $x$? Reading $\BEP^\pm$ as the primary object is what separates frequency of
mismatch (the level near $x = 0$), severity of mismatch (how slowly the curve decays for large
$x$), and direction (shortfall versus overpayment), and it is what lets two candidate contract
shapes be compared even when their AAL is matched---precisely the comparison a single hedge-effectiveness
number cannot support.

\paragraph{Return-period interpretation (basis return periods).} Just as EP curves are often
communicated via return periods, basis EP curves can be summarised by basis return-period levels. For a return period $T$ (years), define the shortfall-basis return-period level
$b_T^+$ by $\BEP^+(b_T^+) = \PR(B > b_T^+) = 1/T$,
and similarly define the overpayment-basis return-period level $b_T^-$ by $\BEP^-(b_T^-) = \PR(-B > b_T^-) = 1/T$.
Reporting $(b_T^+, b_T^-)$ for a small set of $T$ values (e.g.\ $T \in \{10, 20, 50\}$) yields a compact
and interpretable summary of basis-tail behaviour that can sit next to the standard loss
return-period table.

\subsubsection{Portfolio basis: aggregate and occurrence}
Let $B_r = Y_r - P_r$ denote the basis for the risk $r \in \R$. In a multi-region portfolio, basis
reporting must distinguish between (i) the aggregate mismatch over all regions in the model
year, and (ii) the occurrence (worst-region) mismatch that can dominate operational and
regulatory concerns. Given a portfolio, define the aggregate basis as
\[
S^B = \sum_{r\in\R} B_r = S^Y - S^P.
\]
This quantity is natural when the objective is balance-sheet impact and annual aggregate
liquidity, and it aligns with AEP-style reporting for liabilities. For basis regulation and
suitability, we additionally introduce an occurrence basis that captures the worst regional
mismatch within the year:
\[
M^B = \max_{r\in\R} B_r.
\]
The occurrence basis is the appropriate object when basis risk is interpreted as an operational shortfall/overpayment exposure that can be driven by a single region, even if
the portfolio aggregate is moderate. This parallels the NatCat distinction between AEP
and OEP, and it is particularly relevant for supervisory review because it measures basis
accumulation: the most adverse mismatch that the index can generate in a given year.
The associated tail measures are
\[
\BAEP^\pm(x) := \PR(\pm S^B > x), \qquad \BOEP^\pm(x) := \PR(\pm M^B > x),
\]
with basis return levels defined analogously by $\BAEP^\pm(b_T^\pm) = 1/T$ and $\BOEP^\pm(b_T^\pm) = 1/T$.

\subsection{Minimal reporting template}
For each NatPar candidate design (within a region/segment/etc), report:
\begin{enumerate}[leftmargin=1.6em]
\item \textbf{Mean comparability:} AAL as the calibration target.
\item \textbf{Tail comparability:} EP/AEP/OEP, return-period levels.
\item \textbf{Basis interface (central object):} the basis exceedance distribution $\BEP^\pm$ (and portfolio $\BAEP^\pm/\BOEP^\pm$) with return-period levels, reported with the same prominence as the loss EP curve; point diagnostics are read off it.
\item \textbf{Capital metrics:} $\VaR_{0.995}$ (and TVaR where feasible).
\item \textbf{Pricing:} the trigger price from the loss-minimisation of Section~\ref{sec:pricing}, with the chosen penalty weight and curvature reported alongside it.
\end{enumerate}
As we will see in Section~\ref{sec:frost} we implement this reporting template in a frost case study with three simulated regions sharing one identical damage function. We compare a continuous bounded NatPar design to an indemnity benchmark
and then compare an extreme indemnity layer to a binary design, illustrating how payout
shape can dominate tail capital even under AAL-neutral calibration.

\subsection{How the workflow adjusts to basis-risk distribution analysis}\label{sec:adjust}
Because the distributional view of basis risk is the one genuine departure from NatCat practice,
it is worth being explicit about how the three operational functions that consume catastrophe
output---reporting, underwriting, and portfolio management---change once basis risk is reported as
a distribution rather than as a scalar. In each case the existing workflow is retained; what is
added is a basis-distribution analogue of a step the function already performs.

\paragraph{Reporting.} The loss report gains a basis report of equal standing. The organising
principle is a one-to-one correspondence: every loss-side object acquires a basis-side counterpart
reported with the same prominence. The loss EP curve is paired with the two-sided $\BEP^\pm$ curves,
the loss return-period table with the basis return-period levels $(b_T^+, b_T^-)$, and---at
portfolio level---the AEP/OEP curves with the $\BAEP^\pm$ and $\BOEP^\pm$ curves. The $\BEP^\pm$
curve is not an appendix to the loss EP curve but its peer: it is the canonical view of the residual
risk, and the scalar diagnostics (correlation, average shortfall, hedge effectiveness) are summaries
read off it rather than substitutes for it. The reporting unit is unchanged
(region, segment, treaty), and AAL-neutral calibration is the comparability convention that lets
two candidate designs be placed side by side, since it strips out mean-level differences and leaves
only the shape of the mismatch. The practical effect is that a basis report is no longer a single
hedge-effectiveness number that can be gamed by averaging; it exposes whether residual risk is
frequent-and-small or rare-and-severe, and in which direction (shortfall or overpayment), in the
same language the loss report already uses.

\paragraph{Underwriting.} Underwriting acceptance criteria, today expressed as thresholds on loss
metrics, gain basis-distribution thresholds. Rather than accepting a design because a correlation or
hedge-effectiveness ratio clears a cutoff, the underwriter sets limits on the basis tail directly:
a maximum admissible shortfall return level $b_T^+$ at a chosen horizon, a cap on the overpayment
return level $b_T^-$ (which is a cost to the writer, not the client), and a sign/severity profile
that the contract must respect. Trigger and scale are then chosen by the loss-minimisation of
Section~\ref{sec:pricing} subject to those basis-tail limits, so the curvature of the pricing
penalty becomes an explicit underwriting lever: tightening the admissible shortfall tail is the same
operation as raising the curvature on the shortfall side. This also disciplines product complexity,
because an additional trigger or index is justified only when it produces a measurable reduction in
the basis tail net of the transparency and operational robustness it costs.

\paragraph{Portfolio management.} At the book level the manager already monitors loss accumulation
through AEP and OEP; the basis-distribution analogue is to monitor \emph{basis} accumulation through
$S^B$ and $M^B$. The aggregate basis $S^B$ governs annual balance-sheet and liquidity impact, while
the occurrence basis $M^B$---the worst single-region mismatch in the year---is the object that
drives suitability and conduct exposure, because a portfolio whose aggregate basis nets out can
still leave one region badly under-covered. Crucially, as developed in Section~\ref{sec:whogains},
the basis tail of a book is not the sum of its parts: it depends on how the underlying hazards
co-move, and specifically on their \emph{tail dependence} rather than their correlation. Portfolio
construction therefore acquires a basis-tail dimension---diversifying across weakly tail-dependent
hazards genuinely thins the aggregate basis tail, whereas adding exposures whose extremes co-occur
does not---and the manager's stress tests are best specified as joint-extreme (tail-dependence)
scenarios on $M^B$ rather than as correlation matrices on losses.

\section{Pricing the trigger by minimising a loss of mismatch}\label{sec:pricing}
The AAL-neutral calibration of Section~\ref{sec:natpar} fixes the scale of a payout once its
shape and trigger are chosen. In this section we promote the trigger to a \emph{decision
object} and pose pricing as a single optimisation: choose the contract parameters to
minimise the expected disutility of the two-sided mismatch between the parametric payout
and the loss it hedges. The shape of that disutility is controlled by the analyst through
\emph{both} a weight on each side of the basis \emph{and} a curvature; one may also replace
the whole penalty by a general loss function. A particular boundary case of this
problem---when the penalty is symmetric and linear---coincides with minimising the mean
basis and hence with AAL-neutrality; we record that coincidence as a remark rather than as
a separate pricing paradigm.

\subsection{Model and notation}
Let $X$ denote the parametric index (here the damage proxy $D$ derived from the hazard
index $T$), let $L(X)$ be the true loss the buyer wants to hedge (the NatCat damage
loss), and let $P(X;\theta)$ be the parametric payout with contract parameters $\theta$. The
basis is $B = P - L$, with under-pay (shortfall) $B^- = (L-P)^+$ and over-pay
$B^+ = (P-L)^+$; $\AAL = \E[L]$. In the optimisation the loss curve is fixed
exogenously: its anchors are set from index quantiles and its curvature is given, so $L$ is
ground truth, not a decision variable.

For the linear-ramp / exceedance contract the payout is
\[
P(x;s,e) = N\cdot \mathrm{clip}\!\Big(\tfrac{x-s}{e-s}, 0, 1\Big),
\]
i.e.\ zero below strike $s$, ramping to the full notional $N$ at exit $e$. The decision
variables are the two triggers $\theta=(s,e)$. The binary payout uses a single trigger,
$P=N\cdot\mathbf{1}\{x\ge s\}$, hence one unknown. In the frost case study of
Section~\ref{sec:frost} the same logic is carried by the continuous design
$P^c = q^c(D-d)^+$ and the binary design $P^b=q^b\mathbf{1}\{D>d\}$, whose decision
variables are $\theta=(d,q)$ and $\theta=(d)$ respectively (with $q$ pinned by AAL-neutrality
in the single-trigger case).

\subsection{A general loss-minimisation objective}
Let $\rho$ be a disutility of mismatch built from the two nonnegative sides of the basis, the
shortfall $B^-=(L-P)^+$ and the over-pay $B^+=(P-L)^+$. We price by
\begin{equation}\label{eq:objective}
\min_{\theta}\;\; \rho\big(L,P(\theta)\big)
\;:=\;
\lambda\,\E\big[\phi_-\!\big((L-P)^+\big)\big]
\;+\;(1-\lambda)\,\E\big[\phi_+\!\big((P-L)^+\big)\big],
\end{equation}
where $\theta$ are the contract parameters (the triggers and scale), $\lambda\in[0,1]$ weights
shortfall against over-pay, and $\phi_-,\phi_+:\mathbb{R}_+\to\mathbb{R}_+$ are increasing
penalties applied to each side. The analyst shapes the price through two distinct levers:
\begin{itemize}[leftmargin=1.4em]
\item \textbf{Weight} ($\lambda$): a linear tilt that makes shortfall more or less costly than
over-pay without changing how marginal cost grows with the size of the miss.
\item \textbf{Curvature} ($\phi_\pm$): how fast the penalty accelerates in the size of the miss. A
convex $\phi$ penalises large misses disproportionately, so it reshapes the \emph{tail} of the
basis, not merely its average direction.
\end{itemize}
The leading parametric family is the power penalty $\phi_\pm(z)=z^{p_\pm}$ with exponents
$p_-,p_+\ge 1$, which nests the symmetric $L_1$ case $(\lambda,p_-,p_+)=(\tfrac12,1,1)$, the
shortfall-tilted $L_1$ case $(\lambda>\tfrac12,1,1)$, and asymmetric-curvature cases such as
$(p_-,p_+)=(2,1)$ or $(3,1)$ that punish under-payment ever more steeply. Nothing in the
formulation requires a power penalty: $\phi_\pm$ may be any increasing loss---for instance a
Huber penalty (quadratic near zero, linear in the tail), an exponential disutility
$\phi_-(z)=e^{\kappa z}-1$ that targets extreme shortfall, or a log-barrier enforcing a hard cap.
Equation~\eqref{eq:objective} should therefore be read as a general loss-minimisation, with
the power family used only as a transparent default.

In practice we add a soft budget term $\mu\,|\E[P]-\AAL|$ to keep the premium economically
anchored unless the curvature deliberately pulls it away; $\mu$ is a Lagrange-style weight, not
a separate pricing mode. The problem is solved by a grid over the index quantiles followed by
a Nelder--Mead refinement, with expectations evaluated on a Monte-Carlo--simulated index
under common random numbers.

\begin{remark}[The symmetric-linear case is AAL-neutral pricing]\label{rem:fair}
For any real $a$, $a^+-(-a)^+=a$, so
$\E[(P-L)^+]-\E[(L-P)^+]=\E[P]-\AAL$. When the penalty is symmetric and linear
$(\lambda,p_-,p_+)=(\tfrac12,1,1)$, the gradient of $\rho$ in the scale direction vanishes
precisely where $\E[(P-L)^+]=\E[(L-P)^+]$, i.e.\ where $\E[P]=\AAL$. In that boundary case
minimising the mismatch loss \emph{is} AAL-neutral (``fair'') pricing: equal expected shortfall
and over-pay, zero mean basis, and pure-premium scale are one and the same condition.
This is why ``fair'' is not a competing paradigm but the symmetric-linear corner of
\eqref{eq:objective}; any asymmetry in weight or curvature moves the price off that corner in a
controlled, reportable way. The identity is robust to a payout cap, since $a^+-(-a)^+=a$ holds
pointwise for the capped $P$ as well.
\end{remark}

\subsection{Determinacy: parameters, conditions, and well-posedness}
The loss-minimisation~\eqref{eq:objective} has as many free parameters as the payout
provides: two for the linear ramp $(s,e)$ (equivalently $(d,q)$ for the continuous frost
design), one for the binary trigger. How many \emph{exact} averaging conditions a price can
satisfy is bounded by that count. The symmetric-linear corner imposes one condition
($\E[P]=\AAL$, Remark~\ref{rem:fair}); a two-parameter contract therefore retains one degree
of freedom, which any asymmetry in weight or curvature spends on shaping the basis:
\begin{equation}\label{eq:wellposed}
\min_{(s,e)} \;\lambda\,\E[\phi_-((L-P)^+)]+(1-\lambda)\,\E[\phi_+((P-L)^+)]
\qquad(\text{soft budget }\mu\,|\E[P]-\AAL|).
\end{equation}
Generically this has an isolated (locally unique) solution. The single-trigger binary contract
has no spare degree of freedom: the budget alone pins it, so curvature and weight cannot move
its price---they can only be reported as the residual basis they leave behind, which is why in
the empirical results curvature reshapes the continuous price but not the binary one. The
problem becomes over-determined only if one demands more \emph{independent} exact conditions
than parameters: with two triggers a second exact condition (e.g.\ matching loss
$\TVaR_{0.95}$) uses the last degree of freedom, and a third leaves no exact solution. The
design rule is that exactly-satisfiable conditions must not exceed free parameters; each added
parameter (a free cap, a multi-layer attachment, an extra trigger) buys one more, which is also
what lets a richer penalty curvature express itself.

Two well-posedness points complete the picture. First, existence: $\E[P(\theta)]$ is continuous
and monotone in the triggers and ranges in $[0, N\cdot\PR(\text{trigger})]$, so an AAL-neutral
solution exists iff AAL lies in that range; if even the most generous admissible payout has
$\E[P]<\AAL$ (an under-capitalised or capped contract) the budget is infeasible, and the soft
penalty pushing to the boundary is detected and flagged as a signal to re-parameterise---raise
the notional or add a layer---rather than sit silently on a boundary. Second, the objective is
non-convex (a clip composed with the penalty), so multiple local minima are possible under
strong curvature; we therefore solve by a grid followed by Nelder--Mead refinement and report
local solutions. Finally, the objective prices at the physical AAL (pure premium); a risk load
attaches as $\text{premium}=\AAL+\text{load}$ without disturbing the trigger geometry, since it
shifts the budget target rather than the shape of $P$---equivalently, a convex penalty curvature
already encodes a risk attitude endogenously.

\section{Two design blocks and one reporting language for individual risk}\label{sec:blocks}
This section operationalises the NatCat-to-NatPar mapping in a worked frost example on real
data and reports outputs in the same portfolio
language used in catastrophe practice for an individual risk. The point is not to build the
most realistic frost model, but to make the workflow explicit: starting from a NatCat loss
specification, constructing a NatPar payout that is legible to underwriting and governance,
calibrating it to a transparent target, then constructing the reporting framework and finally
comparing the two using the same diagnostics. To better focus on the
reporting and regulatory framework, we assume there is one season per region, so that within a region
$\AEP = \OEP = \EP$. In the numerical assessment we will use a portfolio
in addition to the individual risk.

\subsection{Common notations}
We index regions by $r \in \{\text{R1}, \text{R2}, \text{R3}\}$. In the frost case study each season produces at most
one relevant frost outcome per region. Hence, within a region, annual aggregate and annual
occurrence losses coincide, and the usual catastrophe reporting objects (AEP/OEP) reduce
to a single exceedance curve. This ``one-loss-per-year'' structure is deliberate: it isolates
the effect of contract design and basis risk without confounding from within-year event
counts.

\paragraph{State variables and loss components.} For each region $r$:
\begin{itemize}[leftmargin=1.4em]
\item $T_r$ denotes the (seasonal) trigger, e.g., minimum temperature as hazard proxy/index.
\item $A_r$ denotes seasonal exposure, e.g., monetary value at risk.
\item $D(\tau) \in [0, 1]$ denotes a non-increasing hazard-driven damage fraction (vulnerability
curve). We assume $D(\tau) > 0 \Leftrightarrow \tau < \tau_t$, for a given $\tau_t$, and $D(\tau) < 1 \Leftrightarrow \tau > \tau_c$. The
left inverse of $D$ is denoted by $D^-$.
\item $L_r$ denotes ground-up loss. In this paper we consider the multiplicative form $L_r = A_r D_r$. We use the notation $D_r = D(T_r)$ throughout.
\end{itemize}
Let us introduce the following exposure functions: the CDF $F_A(x) := \PR(A \le x)$, and
the call functional $C_A(x) := \E[(A - x)^+]$.

\paragraph{NatCat vs.\ NatPar payoffs.} We write contractual (NatCat-type) indemnity loss as a
transformation of ground-up loss, $Y_r := \psi(L_r)$, where $\psi$ captures financial terms
(e.g.\ deductible/stop-loss layer). The parametric (NatPar-type) payout depends only on the
hazard proxy, $P_r := I(T_r)$, where $I(\cdot)$ is the payout function (continuous, piecewise, or
binary).

\paragraph{Calibration convention.} Within each region and design
block, we calibrate NatPar by an AAL-neutral principle, $\E[P_r] = \E[Y_r]$,
so differences in EP curves, tail metrics, and basis risk reflect distributional shape and index
mismatch rather than differences in mean payout. This is a comparability device, not a
claim of optimality. In Section~\ref{sec:frost} we additionally report trigger prices from the
loss-minimisation of Section~\ref{sec:pricing} under a panel of penalty shapes.

\paragraph{Ground-up loss.} The seasonal ground-up loss is multiplicative:
$L_r := A_r D(T_r) = A_r D_r$. This is the canonical NatCat construction: hazard produces a
state ($T$), vulnerability maps hazard to a fractional impact ($D$), and exposure scales
impact into monetary loss ($A \times D$). In this form, uncertainty in $A$ loads directly into
the loss distribution whenever $D_r > 0$.

\paragraph{NatCat indemnity cover with deductible.} We benchmark against a simple deductible (stop-loss) indemnity form with region-specific deductible $l_r \ge 0$:
\[
Y_r := (L_r - l_r)^+ = \max(A_r D_r - l_r, 0).
\]
Because the frost case study produces at most one relevant frost loss per season in each
region, $Y_r$ is simultaneously the event loss, the annual occurrence loss, and the annual
aggregate loss for that region.

\subsection{NatCat: analytic EP and AAL representations}
Fix a region $r$ and a threshold $x \ge 0$. The exceedance probability of indemnity loss is
\begin{equation}\label{eq:catep}
\EP^{\Cat}_r(x) = \EP_r(x) := \PR(Y_r > x) = \E\!\left[\overline{F}_{A_r}\!\left(\frac{l_r + x}{D_r}\right)\right],
\end{equation}
where $\overline{F}_{A_r}$ is the exposure survival in region $r$. This form makes explicit the supply-side sensitivity: tail exceedance is controlled by the probability that exposure exceeds a
hazard-dependent threshold proportional to $(l_r+x)/D_r$. Similarly, the average annual loss (AAL) of the deductible indemnity admits the conditional form
\begin{equation}\label{eq:cataal}
\AAL^{\Cat}_r = \E\!\left[D_r\, C_{A_r}\!\left(\frac{l_r}{D_r}\right)\right].
\end{equation}
Equations~\eqref{eq:catep}--\eqref{eq:cataal} show that, in the ground-up NatCat benchmark, exposure uncertainty enters the liability distribution through $F_{A_r}$. This
is the structural mechanism that NatPar will later modify: once payout depends only on $T_r$
(and is AAL-calibrated), exposure uncertainty drops out of the insurer's liability distribution, but it reappears as a measurable residual mismatch in the basis variable $B_r = Y_r - P_r$.

\subsection{NatPar: analytic EP and AAL representations}
The key structural change
relative to the NatCat benchmark is that the payout depends only on the observable hazard
proxy $T_r$ (through the damage function), not on exposure $A_r$. The ground-up payouts are
\[
L^c_r = (D_r - d_r)^+, \qquad L^b_r = \mathbf{1}\{T_r < \tau_r\} = \mathbf{1}\{D_r > d_r\},
\]
where $D(\cdot)$ is the same vulnerability curve used in the NatCat benchmark and $d_r = D(\tau_r)$,
with $\tau_r$ a region-specific trigger. This payoff is (i) fully index-driven, (ii) bounded by
construction, and (iii) interpretable.

\subsubsection{Continuous parametric payout}
Define a continuous payout NatPar as
\begin{equation}\label{eq:contpay}
P^c_r := q^c_r L^c_r = q^c_r (D_r - d_r)^+.
\end{equation}
To compare distributional shape rather than mean level, we choose $q^c_r$ such that NatPar is
AAL-neutral to the NatCat benchmark indemnity $Y_r$:
\begin{equation}\label{eq:contaal}
\AAL_r := \E[P^c_r] = \E[Y_r].
\end{equation}
Substituting~\eqref{eq:contpay} into~\eqref{eq:contaal} yields the explicit calibration
\begin{equation}\label{eq:contq}
q^c_r = \frac{\E[(A_r D_r - l_r)^+]}{\E[(D_r - d_r)^+]} = \frac{\E\!\left[D_r C_{A}\!\left(\frac{l_r}{D_r}\right)\right]}{C_{D_r}(d_r)}.
\end{equation}
Because $P^c_r$ depends only on $T_r$, its EP curve has a simple hazard-only representation:
for any threshold $x \ge 0$,
\begin{equation}\label{eq:contep}
\EP^c_r(x) := \PR(P^c_r > x) = \overline{F}_{D_r}\!\left(\frac{x}{q^c_r} + d_r\right).
\end{equation}

\subsubsection{Binary payout}
The NatPar binary payout is
\begin{equation}\label{eq:binpay}
P^b_r = q^b_r L^b_r = q^b_r \mathbf{1}\{T_r < \tau_r\},
\end{equation}
with AAL-neutral calibration $\E[P^b_r] = \E[Y_r]$. Substituting~\eqref{eq:binpay} into the AAL-neutrality condition gives
\begin{equation}\label{eq:binq}
q^b_r = \frac{\E[(A_r D_r - l_r)^+]}{\PR\{D_r > d_r\}} = \frac{\E\!\left[D_r C_{A}\!\left(\frac{l_r}{D_r}\right)\right]}{\overline{F}_{D_r}(d_r)}.
\end{equation}
The exceedance follows as
\[
\EP^b_r(x) = \begin{cases} \overline{F}_{D_r}(d_r), & x \le q^b_r, \\ 0, & x > q^b_r. \end{cases}
\]

\subsection{Basis risk diagnosis}
Let $B^{\{\cdot\}}_r := Y_r - P^{\{\cdot\}}_r$ for $\cdot \in \{c, b\}$.

\subsubsection{Point diagnosis}
It is not difficult to show that
\[
\E[B^{\{\cdot\},+}_r] = \E\!\left[D_r C_A\!\left(\frac{l_r + P^{\{\cdot\}}_r}{D_r}\right)\right].
\]
Given AAL-neutrality, $\E[B^{\{\cdot\}}_r] = 0$, which implies $\E[B^{\{\cdot\},+}_r] = \E[B^{\{\cdot\},-}_r]$.
In addition,
\[
\Cov(Y_r, P^{\{\cdot\}}_r) = \E\!\left[P^{\{\cdot\}}_r D_r C_A\!\left(\frac{l_r}{D_r}\right)\right] - (\AAL^{\{\cdot\}}_r)^2.
\]

\paragraph{Continuous NatPar.} With $P^c_r$:
\begin{align*}
\E[B^{c,+}_r] &= \E\!\left[D_r C_{A_r}\!\left(\tfrac{l_r}{D_r}\right)\mathbf{1}\{D_r\le d_r\}\right] + \E\!\left[D_r C_{A_r}\!\left(\tfrac{l_r}{D_r}+q^c_r\big(1-\tfrac{d_r}{D_r}\big)\right)\mathbf{1}\{D_r>d_r\}\right], \\
\Var(P^c_r) &= (q^c_r)^2\!\left(\E\!\left[D_r^2\big(1-\tfrac{d_r}{D_r}\big)_+^2\right] - \E\!\left[D_r\big(1-\tfrac{d_r}{D_r}\big)_+\right]^2\right), \\
\Cov(Y_r, P^c_r) &= q^c_r\,\E\!\left[D_r^2\big(1-\tfrac{d_r}{D_r}\big)_+^2 C_A\!\left(\tfrac{l_r}{D_r}\right)\right] - (\AAL_r)^2.
\end{align*}

\paragraph{Binary NatPar.} With $P^b_r$:
\begin{align*}
\E[B^{b,+}_r] &= \E\!\left[D_r C_{A_r}\!\left(\tfrac{l_r}{D_r}\right)\mathbf{1}\{D_r\le d_r\}\right] + \E\!\left[D_r C_{A_r}\!\left(\tfrac{l_r+q^b_r}{D_r}\right)\mathbf{1}\{D_r>d_r\}\right], \\
\Var(P^b_r) &= (q^b_r)^2\,\PR(D_r>d_r)(1-\PR(D_r>d_r)), \\
\Cov(Y, P^b) &= q^b_r\!\left(\E\!\left[D_r C_{A_r}\!\left(\tfrac{l_r}{D_r}\right)\right] - \AAL_r\PR(D_r>d_r)\right).
\end{align*}

\subsubsection{Basis exceedance curve}
We treat basis risk as a distributional object via the exceedance-basis curves
$\BEP^{\{\cdot\},+}_r(x) := \PR(B^{\{\cdot\}}_r > x)$ and $\BEP^{\{\cdot\},-}_r(x) := \PR(-B^{\{\cdot\}}_r > x)$ for $x \ge 0$.
Here $B^{\{\cdot\}}_r > 0$ is shortfall and $-B^{\{\cdot\}}_r > 0$ is overpayment. Under AAL-neutrality,
$\E[B^{\{\cdot\}}_r] = 0$, but this does not control tails. Fixing $x\ge 0$ and conditioning on $T_r$:

\paragraph{Shortfall exceedance.} If $D_r=0$, then $Y_r=P^c_r=0$ and $B^c_r=0$. If $D_r>0$,
$B^{\{\cdot\}}_r > x \Leftrightarrow A_r > (l_r+x+P^{\{\cdot\}}_r)/D_r$, hence
\begin{equation}\label{eq:bepshort}
\BEP^{\{\cdot\},+}_r(x) = \E\!\left[\overline{F}_{A_r}\!\left(\frac{l_r+x+P^{\{\cdot\}}_r}{D_r}\right)\right].
\end{equation}

\paragraph{Overpayment exceedance.} Combining the cases $A_r D_r - l_r > 0$ and $A_r D_r - l_r \le 0$,
\begin{equation}\label{eq:bepover}
\BEP^{\{\cdot\},-}_r(x) = \E\!\left[F_{A_r}\!\left(\frac{l_r-x+P^{\{\cdot\}}_r}{D_r}\right)\mathbf{1}\{P^{\{\cdot\}}_r>x\}\right].
\end{equation}

\subsection{Continuous NatPar basis analysis}
Under the baseline independence assumptions:
\begin{align}
\BEP^{c,+}_r(x) &= \E\!\left[\overline{F}_{A_r}\!\left(\tfrac{l_r+x}{D_r}+q^c_r\big(1-\tfrac{d_r}{D_r}\big)\right)\mathbf{1}\{D_r>d_r\}\right] + \E\!\left[\overline{F}_{A_r}\!\left(\tfrac{l_r+x}{D_r}\right)\mathbf{1}\{D_r\le d_r\}\right], \label{eq:contshort}\\
\BEP^{c,-}_r(x) &= \E\!\left[F_{A_r}\!\left(\tfrac{l_r-x}{D_r}+q^c_r\big(1-\tfrac{d_r}{D_r}\big)\right)\mathbf{1}\{D_r q^c_r(1-d_r/D_r)>x\}\right]. \label{eq:contover}
\end{align}

\subsection{Binary NatPar basis analysis}
\begin{align}
\BEP^{b,+}_r(x) &= \E\!\left[\overline{F}_{A_r}\!\left(\tfrac{l_r+x+q^b_r}{D_r}\right)\mathbf{1}\{D_r>d_r\}\right] + \E\!\left[\overline{F}_{A_r}\!\left(\tfrac{l_r+x}{D_r}\right)\mathbf{1}\{D_r\le d_r\}\right], \label{eq:binshort}\\
\BEP^{b,-}_r(x) &= \begin{cases} \E\!\left[F_{A_r}\!\left(\tfrac{l_r-x+q^b_r}{D_r}\right)\mathbf{1}\{D_r>d_r\}\right], & q^b_r > x, \\ 0, & q^b_r \le x. \end{cases} \label{eq:binover}
\end{align}

\section{Reporting the frost risk example}\label{sec:frost}
We illustrate the framework on three regions that share a single, identical frost damage
function---the convex citrus curve of \citet{assa2017frost} with the published parameters---and
differ only in their night-minimum temperature law. Because real gridded temperature series for
arbitrary locations do not, in general, fall within the narrow citrus damaging band on which the
published curve is calibrated, we drive the three regions with \emph{simulated} night-minimum
temperatures, positioned so that the same fixed damage function is exercised with graded severity.
The three regions are ordered by frost exposure---$\text{R1}$ most exposed, $\text{R2}$
intermediate, $\text{R3}$ least---mirroring the way the three months (January, December, February)
of the published San Joaquin study differ in frost risk. The objective of this subsection is
twofold: (i) to make the NatCat hazard--exposure--vulnerability loss construction explicit, and
(ii) to express key distributional objects (EP curves and AAL) in a conditional form that carries
over to basis-risk analysis.

\subsection{Hazard--exposure--vulnerability and NatCat indemnity}
The frost loss model we use builds directly on the indemnity-based frost-pricing framework of
\citet{assa2017frost}. That paper deliberately prices frost as an \emph{indemnity} contract on the
loss---avoiding an index precisely because an index introduces basis risk---using distortion
premiums (VaR, CVaR, Wang). Here we take its loss model as the ground-up NatCat loss and then ask
the complementary, supply-side question: what happens when one \emph{does} make a bounded index
contractual, and reports the resulting basis risk in the NatCat language. The present case study is
thus the parametric counterpart of \citet{assa2017frost}, sharing its loss generating process.

\paragraph{Hazard (temperature).} For each region $r$, the seasonal night-minimum temperature
$T_I$ is modelled as Gaussian with region-specific parameters and independent across seasons. In
the loss model below this $T_I$ plays the role of the night minimum, and the sunset/average
temperature is reconstructed as $T_0=T_I+N$ via the fitted nighttime drop $N$. Since the case
study is simulation-based, the three regions are defined by their night-minimum laws (degrees
Fahrenheit): $\text{R1}\sim\mathcal N(25.5,2.6^2)$, $\text{R2}\sim\mathcal N(26.5,2.5^2)$,
$\text{R3}\sim\mathcal N(27.5,2.4^2)$, positioned around the citrus damaging band so that the
single identical damage function is exercised with graded severity. The nighttime drop is taken
from the published January fit, $N\sim\mathcal N(9.41,1.41^2)$ in degrees Fahrenheit
\citep{assa2017frost}. The Gaussian assumption is used for transparency; any alternative calibrated
marginal can be substituted without changing the NatCat--NatPar logic.

\begin{remark}[Where the catastrophe nature lives, and tail dependence under Gaussian margins]\label{rem:catnature}
The Gaussian hazard is light-tailed, so it carries no catastrophe character on its own. This is
deliberate, and it locates the catastrophe nature of the loss downstream of the index, in two
channels. First, the loss $L_r=A_r D(T_r)$ inherits the \emph{heavy} right tail of the lognormal
exposure $A_r$ whenever $D_r>0$; the unbounded severity that the bounded parametric payout
cannot follow is exposure-driven, not hazard-driven. Second, the vulnerability map $D(\cdot)$ is
sharply convex near the cold tail, so even a symmetric, light-tailed $T_r$ produces a highly
skewed damage variable $D_r$ (a mass near zero plus a spike near one). The framework is
marginal-agnostic: a genuinely heavy-tailed or clustered hazard---a generalised-extreme-value or
generalised-Pareto fit to block minima, a skew-$t$, or a regime-switching process---can replace the
Gaussian marginal without altering any of the reporting or pricing logic, and would simply add a
third, hazard-side source of tail severity.

Crucially, tail dependence is a separate matter from the marginal. By Sklar's theorem the joint law
factors into marginals and a copula, and the coefficient of tail dependence is a property of the
\emph{copula} alone. One may therefore impose joint cold extremes among Gaussian-margined
temperatures without making any single region's temperature heavy-tailed. The one-factor model of
Section~\ref{sec:diversification} couples the Gaussian margins with a \emph{Gaussian} copula, which
has zero tail dependence for every $\rho<1$; the experiment of Section~\ref{sec:taildep} keeps the
same Gaussian margins but replaces the copula with a $t$-copula, so that $\lambda_L>0$ while each
regional temperature remains exactly Gaussian. The deep-tail relocation studied there is thus driven
by the dependence structure, not by the marginal tails, and is well-posed despite the light-tailed
hazard.
\end{remark}

\paragraph{Exposure (value at risk).} Exposure is modelled as lognormal,
$A_r \sim \mathrm{LogNormal}(\mu_{A,r}, \sigma_{A,r}^2)$, $A_r > 0$. Rather than parameterising
by $(\mu_{A,r},\sigma_{A,r})$ directly, we specify by region the mean $\bar A_r := \E[A_r]$ and the
coefficient of variation $\mathrm{CV}_{A,r} := \sqrt{\Var(A_r)}/\bar A_r$, which implies
$\sigma_{A,r}^2 = \ln(1+\mathrm{CV}_{A,r}^2)$ and $\mu_{A,r} = \ln(\bar A_r) - \tfrac12\sigma_{A,r}^2$.
We set $\bar A_r = 100$ for all regions and $\mathrm{CV}_{A,r} = (0.60, 0.45, 0.50)$ for
$(\text{R1},\text{R2},\text{R3})$, encoding region-specific exposure uncertainty. As a
simplifying baseline we assume $T_r \perp A_r$.

\begin{figure}[H]
\centering
\includegraphics[width=0.62\textwidth]{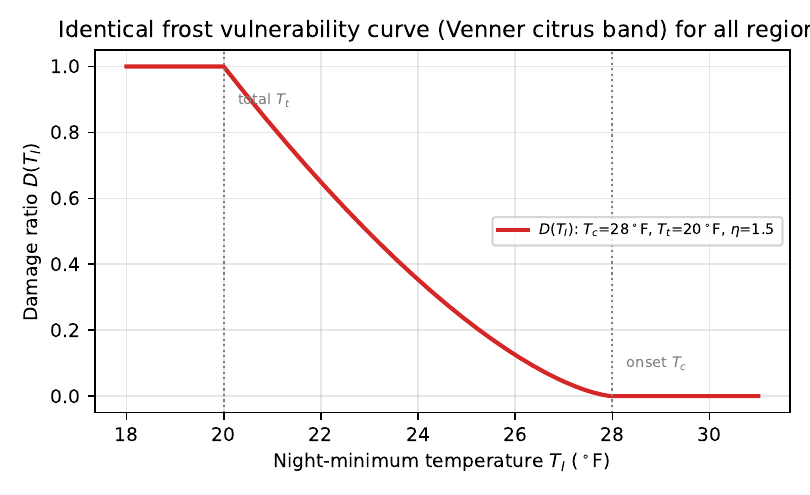}
\caption{The single, identical frost vulnerability curve $D(T_I)$ used for all three regions,
following the \citet{venner1995} convex damage model with the published citrus parameters
$T_c=28^\circ$F, $T_t=20^\circ$F (band width $b=8^\circ$F) and curvature $\eta=1.5$. Damage is
further gated by the \citet{snyder2005} $k$-hour resistance condition $M\le BN$ as in
\citet{assa2017frost}. The regions differ only in their night-minimum temperature law, not in
this curve.}
\label{fig:damage}
\end{figure}

\paragraph{Vulnerability and crop resistance.} For the damage model
we adopt the frost-loss construction of \citet{assa2017frost}, which combines the convex
vulnerability curve of \citet{venner1995} with the square-root nighttime-cooling forecast of
\citet{snyder2005} to capture two features specific to frost: the damage is a \emph{non-linear}
function of temperature, and the crop \emph{resists} the first few hours of freezing. Let $T_I$ be
the night-minimum temperature and $T_0$ the sunset (here, average) temperature. The Venner damage
fraction on the night minimum is
\[
D(T_I) = \begin{cases} 0, & T_I > T_c, \\ \left(\dfrac{T_c - T_I}{T_c - T_t}\right)^{\eta}, & T_t \le T_I \le T_c, \\ 1, & T_I < T_t, \end{cases}
\]
with the citrus parameters $T_c=28^\circ$F, $T_t=20^\circ$F (band width $b=T_c-T_t=8^\circ$F), and
curvature $\eta=1.5$. Crop resistance enters through the requirement that the temperature stay $k$
hours below $T_c$ before damage occurs ($k=4$ for citrus). Under the square-root cooling
$T_i = T_0 + (T_I-T_0)\sqrt{i/I}$ over a night of length $I$ hours, this resistance condition becomes
$M \le B\,N$, where $M=T_0-T_c$, $N=T_0-T_I$ are the (independent) warm-margin and nighttime drop
and $B=\sqrt{(I-k)/I}$. Realised damage is therefore
\[
D = D(T_I)\,\mathbf{1}\{M \le B\,N,\; M \le N \le M+b\},
\]
so damage is active only when the night minimum lies in the damaging band \emph{and} the crop's
$k$-hour resistance is overcome. We follow \citet{assa2017frost} in treating $M$ and $N$ as
independent; the analytic loss CDF of that paper carries over, and we evaluate the resulting loss
distribution by simulation here. The citrus band $(T_c, T_t)$, the curvature $\eta$, the band width
$b$, and the entire resistance/$M$--$N$ machinery are held \emph{identical across all three
regions}---this is exactly the published damage function, unrelocated. The regions differ only in
their night-minimum temperature law, so any difference in their loss distributions is attributable
to the hazard, not to the vulnerability curve. Figure~\ref{fig:damage} shows the single shared
vulnerability curve.

\subsection{Basis risk diagnostics}
This subsection compares the NatCat extreme-layer benchmark to two NatPar designs
(continuous and binary) using exceedance and return-period diagnostics. Within each
$(r,\beta)$ block we price the trigger by the loss-minimisation of Section~\ref{sec:pricing}
under the symmetric-linear penalty, so the price is AAL-neutral and all differences reported
below come from tail shape and basis structure, not from mean level. For each region $r$ we
simulate $N=20{,}000$ seasons and compute damage and loss as $D_r = D(T_r) \in [0,1]$ and
$L_r = A_r D_r$. For each $\beta \in \{0.90, 0.95, 0.99\}$ we define the NatCat extreme-layer
benchmark via $l_{r,\beta} := \VaR_\beta(L_r)$, $Y_{r,\beta} := (L_r - l_{r,\beta})^+$.

\subsubsection{Individual risk}
We evaluate two NatPar archetypes,
$P^c_{r,\beta} = q^c_{r,\beta}(D_r - d^c_{r,\beta})^+$ and $P^b_{r,\beta} = q^b_{r,\beta}\mathbf{1}\{D_r > d^b_{r,\beta}\}$.
Rather than pinning the trigger at an ad-hoc percentile, the triggers $d^{\{\cdot\}}_{r,\beta}$ are the
\emph{optimal} solutions of~\eqref{eq:objective}: for the continuous design the pair
$(d^c_{r,\beta},q^c_{r,\beta})$ is chosen jointly to minimise the symmetric-linear basis loss; for
the binary design the single trigger $d^b_{r,\beta}$ is optimised with $q^b_{r,\beta}$ pinned by
AAL-neutrality. Because the symmetric-linear penalty sits on the AAL-neutral corner
(Remark~\ref{rem:fair}), each design still satisfies $\E[P^{\{\cdot\}}_{r,\beta}]=\E[Y_{r,\beta}]$, so the
comparison remains like-for-like in mean.

\paragraph{Optimal versus quantile triggers.} To show that the optimisation does real work---and
where it does not---Table~\ref{tab:trigcomp} contrasts the optimal trigger with the natural
alternative of pinning the trigger at the loss quantile, $d_{r,\beta}=\VaR_\beta(D_r)$, holding both
designs AAL-neutral so the comparison is like-for-like in mean and only the basis dispersion moves.
Two patterns emerge. For the \emph{continuous} design the optimal trigger sits systematically
\emph{below} the quantile pin and reduces basis variance by roughly $8$--$24\%$, the improvement
growing with $\beta$: pinning the ramp at a high loss percentile starts it too late, leaving
frequent moderate shortfalls that a lower, optimised onset removes. For the \emph{binary} design the
two triggers land close and the optimisation yields essentially no gain (and occasionally a slight
loss in variance), because the single trigger is already pinned by the AAL budget---there is no
genuine free degree of freedom for the loss-minimisation to exploit, exactly as the determinacy
discussion of Section~\ref{sec:pricing} anticipates. The practical reading is that trigger
optimisation matters for graded (continuous) payouts and is close to a no-op for digital ones, so
the value of pricing-as-loss-minimisation is realised precisely where the contract has spare shape
to give.

\begin{table}[H]
\centering
\caption{Optimal trigger versus the quantile pin $d=\VaR_\beta(D)$, both AAL-neutral. Columns: quantile trigger $d_{\mathrm{qtl}}$, optimal trigger $d_{\mathrm{opt}}$, basis variance under each, and the percentage reduction from optimising. The optimisation reduces basis dispersion for the continuous design (c) but is essentially inert for the binary design (b), whose single trigger is pinned by the budget.}
\label{tab:trigcomp}
\begin{tabular}{llcrrrrr}
\toprule
Region & $\beta$ & Design & $d_{\mathrm{qtl}}$ & $d_{\mathrm{opt}}$ & $\Var(B)_{\mathrm{qtl}}$ & $\Var(B)_{\mathrm{opt}}$ & reduction \% \\
\midrule
R1 & 0.90 & c & 0.593 & 0.483 & 238.0 & 198.3 & 16.7 \\
R1 & 0.90 & b & 0.593 & 0.600 & 212.5 & 214.7 & -1.0 \\
R1 & 0.95 & c & 0.720 & 0.591 & 157.2 & 124.0 & 21.1 \\
R1 & 0.95 & b & 0.720 & 0.732 & 137.8 & 140.3 & -1.8 \\
R1 & 0.99 & c & 0.912 & 0.816 & 42.3 & 32.8 & 22.5 \\
R1 & 0.99 & b & 0.912 & 0.862 & 38.8 & 33.7 & 13.1 \\
R2 & 0.90 & c & 0.436 & 0.378 & 85.0 & 74.6 & 12.2 \\
R2 & 0.90 & b & 0.436 & 0.499 & 84.1 & 89.1 & -5.9 \\
R2 & 0.95 & c & 0.557 & 0.488 & 60.9 & 51.4 & 15.6 \\
R2 & 0.95 & b & 0.557 & 0.585 & 53.5 & 56.2 & -5.0 \\
R2 & 0.99 & c & 0.798 & 0.706 & 20.8 & 15.9 & 23.5 \\
R2 & 0.99 & b & 0.798 & 0.800 & 17.5 & 17.5 & -0.3 \\
R3 & 0.90 & c & 0.302 & 0.261 & 67.7 & 62.4 & 7.8 \\
R3 & 0.90 & b & 0.302 & 0.386 & 79.7 & 80.5 & -0.9 \\
R3 & 0.95 & c & 0.418 & 0.370 & 52.2 & 46.1 & 11.7 \\
R3 & 0.95 & b & 0.418 & 0.465 & 50.5 & 52.7 & -4.4 \\
R3 & 0.99 & c & 0.648 & 0.543 & 22.2 & 17.0 & 23.4 \\
R3 & 0.99 & b & 0.648 & 0.623 & 18.0 & 17.6 & 2.2 \\
\bottomrule
\end{tabular}
\end{table}

\begin{figure}[H]
\centering
\includegraphics[width=0.92\textwidth]{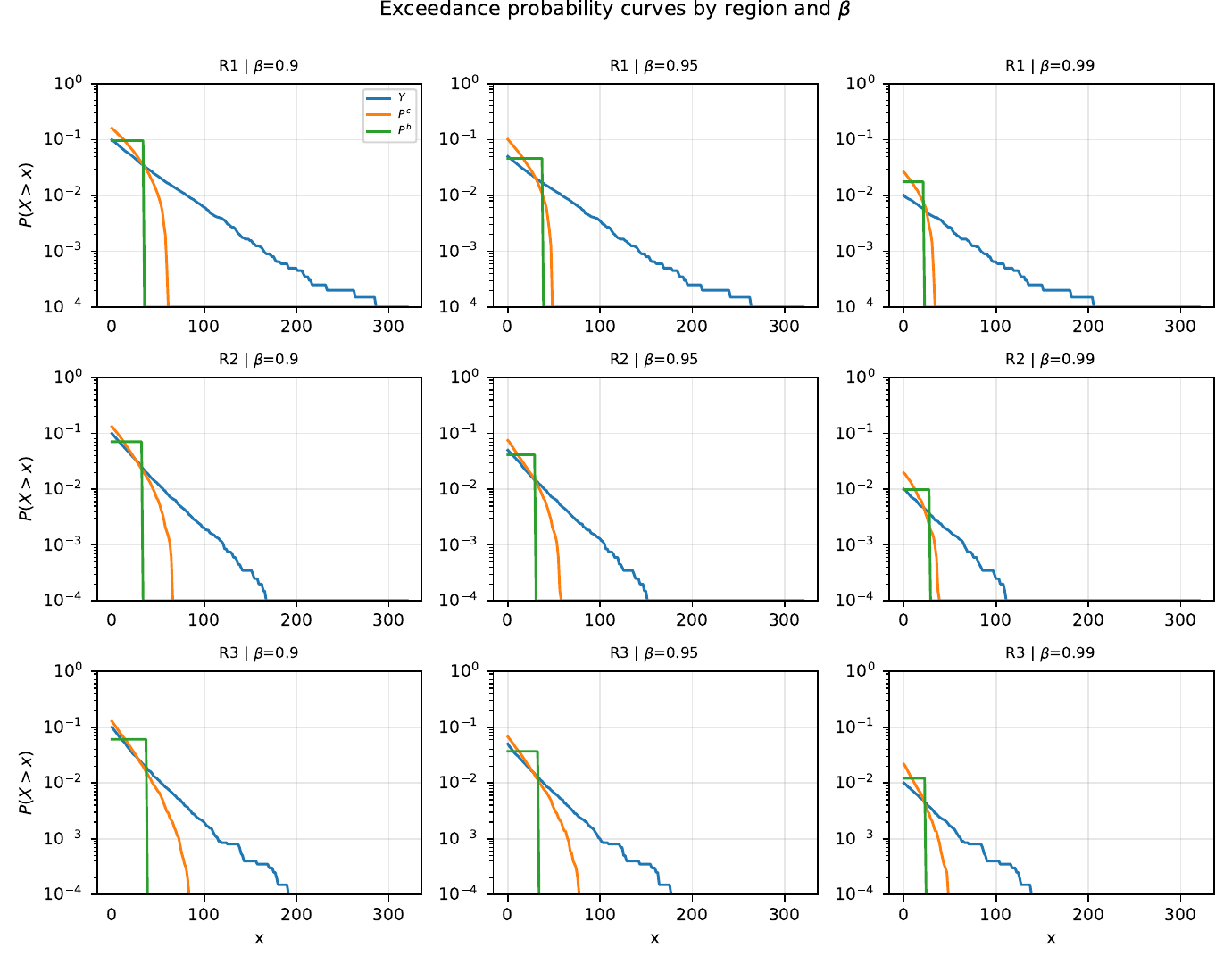}
\caption{Exceedance probability curves for $Y_{r,\beta}$, $P^{\{\cdot\}}_{r,\beta}$ (log scale), by region and $\beta$.}
\label{fig:ep_indiv}
\end{figure}

\paragraph{Exceedance probability curves.} Figure~\ref{fig:ep_indiv} compares the exceedance
probability functions on a log scale. The binary design places a point mass at $q^b_{r,\beta}$, so
$\EP^{b,+}(x)$ is essentially flat up to that level and drops to zero immediately after, creating
the visible ``digital'' tail geometry relative to the smoother benchmark and continuous design.

\paragraph{Point diagnosis of basis risk.} Table~\ref{tab:t1} reports the calibration levels and
dependence diagnostics: the layer attachment $l_{r,\beta}$, the \emph{optimal} damage trigger
$d^{\{\cdot\}}_{r,\beta}$ and scale $q_{r,\beta}$, $\E[Y_{r,\beta}]$, $\Var(B)$, and
$\Corr(Y_{r,\beta}, P^{\{\cdot\}}_{r,\beta})$. Table~\ref{tab:t2} reports the basis mismatch
decomposition: $\PR(B>0)$, $\PR(B<0)$, average shortfall $\E[B^+]$, and average
overpayment $\E[(-B)^+]$.

\begin{table}[H]
\centering
\caption{Point diagnosis: calibration levels and dependence. The trigger $d$ is the optimal solution of the symmetric-linear loss-minimisation; $l_{r,\beta}=\VaR_\beta(L_r)$ is the layer attachment.}
\label{tab:t1}
\begin{tabular}{llcrrrrrr}
\toprule
Region & $\beta$ & Design & $l_{r,\beta}$ & $d_{r,\beta}$ & $q_{r,\beta}$ & $\E[Y_\beta]$ & $\Var(B)$ & $\Corr(Y_\beta,P_\beta)$ \\
\midrule
R1 & 0.90 & c & 59.99 & 0.483 & 116.68 & 3.39 & 198.25 & 0.45 \\
R1 & 0.90 & b & 59.99 & 0.600 & 35.17 & 3.39 & 214.67 & 0.40 \\
R1 & 0.95 & c & 82.60 & 0.591 & 117.67 & 1.77 & 124.01 & 0.34 \\
R1 & 0.95 & b & 82.60 & 0.732 & 38.43 & 1.77 & 140.34 & 0.31 \\
R1 & 0.99 & c & 141.37 & 0.816 & 181.15 & 0.38 & 32.75 & 0.15 \\
R1 & 0.99 & b & 141.37 & 0.862 & 21.48 & 0.38 & 33.71 & 0.13 \\
R2 & 0.90 & c & 42.57 & 0.378 & 106.57 & 2.39 & 74.64 & 0.61 \\
R2 & 0.90 & b & 42.57 & 0.499 & 33.73 & 2.39 & 89.07 & 0.54 \\
R2 & 0.95 & c & 58.46 & 0.488 & 111.08 & 1.25 & 51.40 & 0.49 \\
R2 & 0.95 & b & 58.46 & 0.585 & 30.00 & 1.25 & 56.16 & 0.45 \\
R2 & 0.99 & c & 97.87 & 0.706 & 128.04 & 0.27 & 15.93 & 0.24 \\
R2 & 0.99 & b & 97.87 & 0.800 & 27.86 & 0.27 & 17.52 & 0.22 \\
R3 & 0.90 & c & 28.84 & 0.261 & 113.75 & 2.27 & 62.42 & 0.67 \\
R3 & 0.90 & b & 28.84 & 0.386 & 37.42 & 2.27 & 80.47 & 0.59 \\
R3 & 0.95 & c & 43.72 & 0.370 & 123.35 & 1.21 & 46.06 & 0.57 \\
R3 & 0.95 & b & 43.72 & 0.465 & 32.81 & 1.21 & 52.75 & 0.51 \\
R3 & 0.99 & c & 82.03 & 0.543 & 106.42 & 0.29 & 17.02 & 0.34 \\
R3 & 0.99 & b & 82.03 & 0.623 & 23.93 & 0.29 & 17.58 & 0.32 \\
\bottomrule
\end{tabular}
\end{table}

\begin{table}[H]
\centering
\caption{Point diagnosis: basis mismatch decomposition.}
\label{tab:t2}
\begin{tabular}{llcrrrrr}
\toprule
Region & $\beta$ & Design & $\PR(P>0)$ & $\PR(B>0)$ & $\PR(B<0)$ & $\E[B^+]$ & $\E[(-B)^+]$ \\
\midrule
R1 & 0.90 & c & 0.16 & 0.07 & 0.12 & 2.09 & 2.09 \\
R1 & 0.90 & b & 0.10 & 0.07 & 0.07 & 2.16 & 2.16 \\
R1 & 0.95 & c & 0.10 & 0.04 & 0.08 & 1.28 & 1.28 \\
R1 & 0.95 & b & 0.05 & 0.04 & 0.04 & 1.29 & 1.29 \\
R1 & 0.99 & c & 0.03 & 0.01 & 0.02 & 0.33 & 0.33 \\
R1 & 0.99 & b & 0.02 & 0.01 & 0.02 & 0.34 & 0.34 \\
R2 & 0.90 & c & 0.13 & 0.06 & 0.09 & 1.16 & 1.16 \\
R2 & 0.90 & b & 0.07 & 0.07 & 0.05 & 1.25 & 1.25 \\
R2 & 0.95 & c & 0.08 & 0.03 & 0.06 & 0.74 & 0.74 \\
R2 & 0.95 & b & 0.04 & 0.04 & 0.03 & 0.77 & 0.77 \\
R2 & 0.99 & c & 0.02 & 0.01 & 0.02 & 0.22 & 0.22 \\
R2 & 0.99 & b & 0.01 & 0.01 & 0.01 & 0.22 & 0.22 \\
R3 & 0.90 & c & 0.13 & 0.06 & 0.09 & 1.02 & 1.02 \\
R3 & 0.90 & b & 0.06 & 0.07 & 0.04 & 1.13 & 1.13 \\
R3 & 0.95 & c & 0.07 & 0.03 & 0.05 & 0.66 & 0.66 \\
R3 & 0.95 & b & 0.04 & 0.03 & 0.03 & 0.69 & 0.69 \\
R3 & 0.99 & c & 0.02 & 0.01 & 0.02 & 0.22 & 0.22 \\
R3 & 0.99 & b & 0.01 & 0.01 & 0.01 & 0.22 & 0.22 \\
\bottomrule
\end{tabular}
\end{table}

\begin{figure}[H]
\centering
\includegraphics[width=0.92\textwidth]{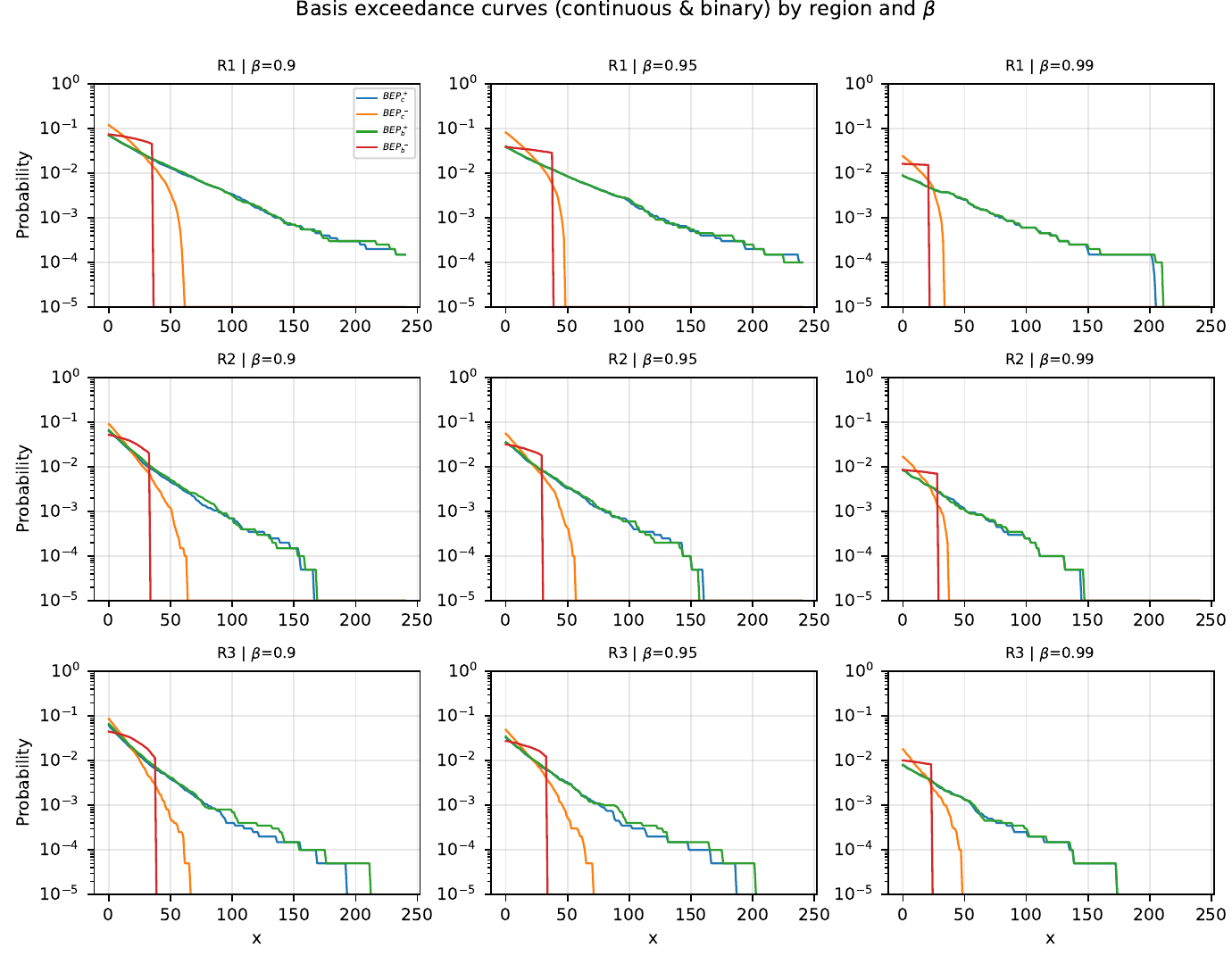}
\caption{Basis exceedance curves for continuous and binary NatPar by region and $\beta$ (log scale).}
\label{fig:basis_indiv}
\end{figure}

\paragraph{Basis exceedance curves.} Figure~\ref{fig:basis_indiv} reports the two-sided basis
exceedance functions, separating consumer-protection tail risk (shortfall) from
capital/pricing tail risk (overpayment). The binary design exhibits a sharper truncation in
the overpayment tail (due to the fixed payout size), while the continuous design decays more
smoothly.

\begin{figure}[H]
\centering
\includegraphics[width=0.92\textwidth]{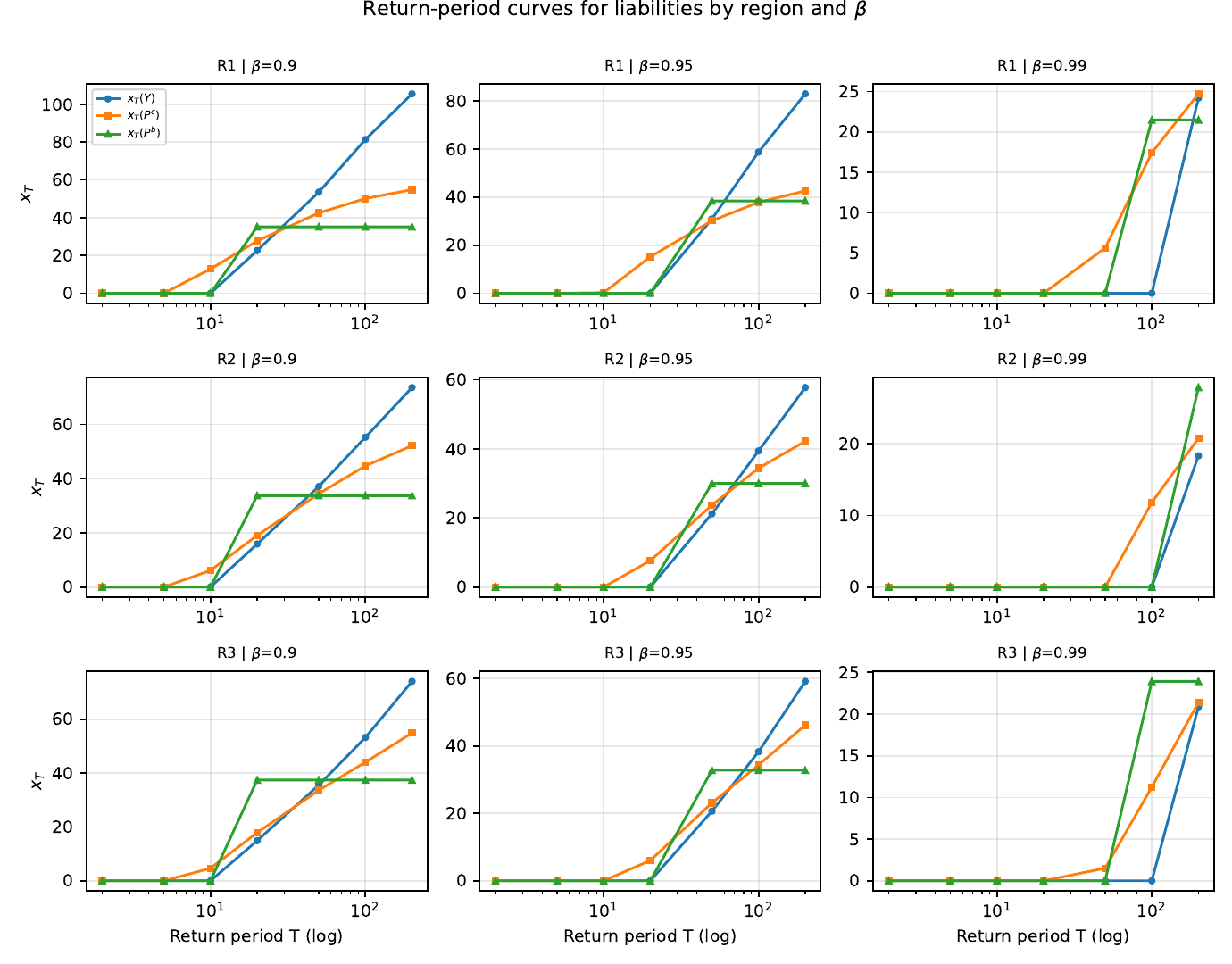}
\caption{Return-period curves $T\mapsto x_T(\cdot)$ for $Y_{r,\beta}$, $P^{\{\cdot\}}_{r,\beta}$, by region and $\beta$.}
\label{fig:rp_liab}
\end{figure}

\begin{figure}[H]
\centering
\includegraphics[width=0.92\textwidth]{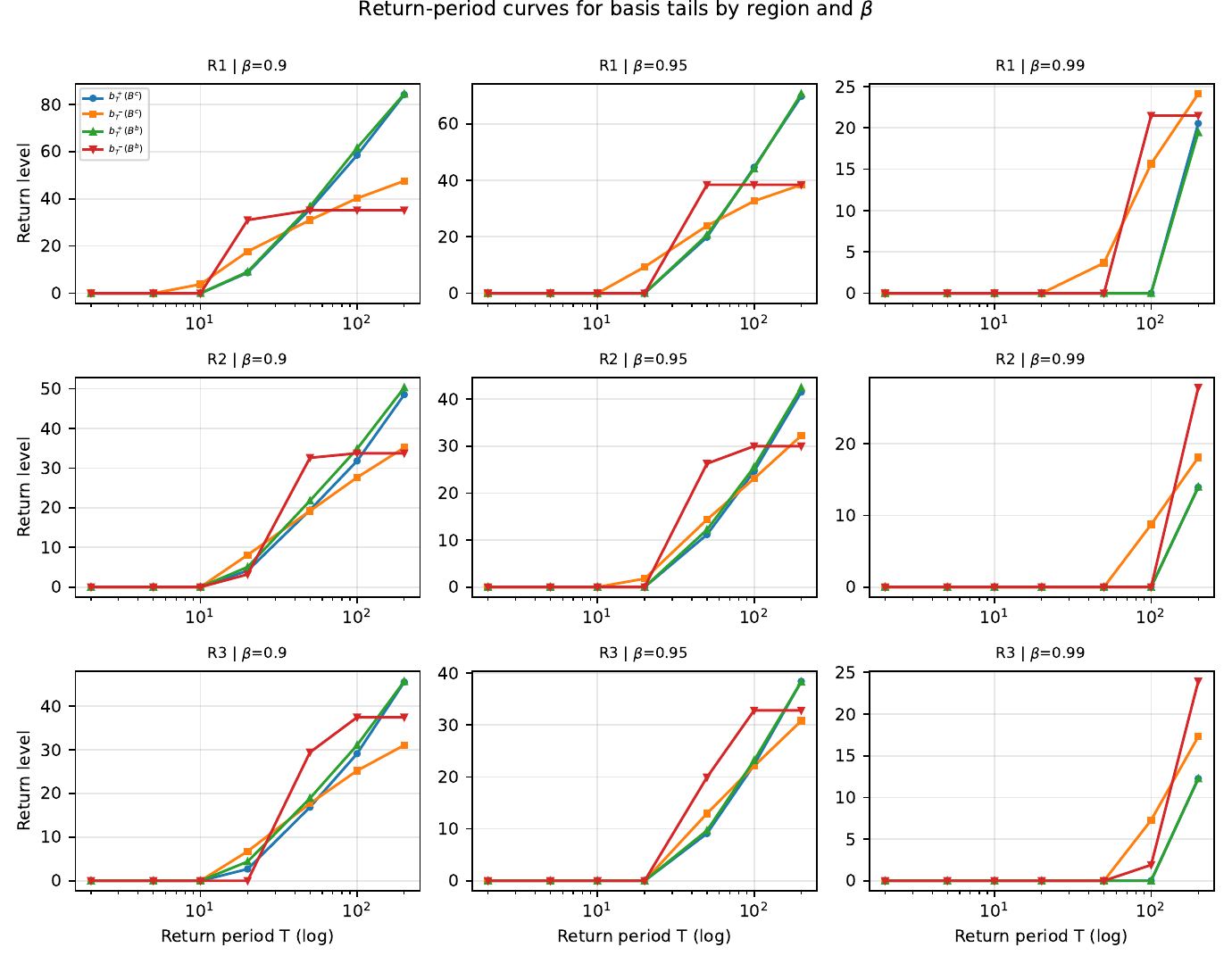}
\caption{Return-period curves for basis tails: $T\mapsto b^+_T$ (shortfall) and $T\mapsto b^-_T$ (overpayment), for both continuous and binary NatPar, by region and $\beta$.}
\label{fig:rp_basis}
\end{figure}

\paragraph{Return-period levels.} A key feature in the results is the presence of many zeros at
low return periods: for the layer $Y_{r,\beta} = (L - l_{r,\beta})^+$, we have $\PR(Y_{r,\beta}>0)=1-\beta$,
so when $1/T \ge 1-\beta$ the return level is exactly zero. The binary payout also produces
plateau behaviour in $x_T(P^b_{r,\beta})$ because of its point mass at $q^b_{r,\beta}$; see
Figures~\ref{fig:rp_liab} and~\ref{fig:rp_basis}.

\paragraph{Capital metrics.} Table~\ref{tab:t3} reports $\VaR_{0.995}$ and $\TVaR_{0.995}$ for
$Y_{r,\beta}$ and each NatPar design across $\beta$. In this case-study setting (at most one
seasonal event per year) these correspond directly to EP/AEP return levels at exceedance
probability $0.5\%$.

\begin{table}[H]
\centering
\caption{Capital metrics for NatCat layer $Y_{r,\beta}$ and NatPar liabilities $P^{\{\cdot\}}_{r,\beta}$: $\VaR_{0.995}$ and $\TVaR_{0.995}$, by region and layer $\beta$.}
\label{tab:t3}
\begin{tabular}{llrrrrrr}
\toprule
Region & $\beta$ & $\VaR(Y_\beta)$ & $\TVaR(Y_\beta)$ & $\VaR(P^c_\beta)$ & $\TVaR(P^c_\beta)$ & $\VaR(P^b_\beta)$ & $\TVaR(P^b_\beta)$ \\
\midrule
R1 & 0.90 & 105.58 & 145.35 & 54.84 & 57.29 & 35.17 & 35.17 \\
R1 & 0.95 & 82.97 & 122.75 & 42.59 & 45.06 & 38.43 & 38.43 \\
R1 & 0.99 & 24.20 & 63.97 & 24.72 & 28.52 & 21.48 & 21.48 \\
R2 & 0.90 & 73.63 & 101.91 & 52.19 & 58.09 & 33.73 & 33.73 \\
R2 & 0.95 & 57.75 & 86.02 & 42.21 & 48.35 & 30.00 & 30.00 \\
R2 & 0.99 & 18.33 & 46.61 & 20.78 & 27.87 & 27.86 & 27.86 \\
R3 & 0.90 & 74.06 & 102.57 & 54.89 & 64.53 & 37.42 & 37.42 \\
R3 & 0.95 & 59.18 & 87.69 & 46.17 & 56.62 & 32.81 & 32.81 \\
R3 & 0.99 & 20.87 & 49.37 & 21.41 & 30.42 & 23.93 & 23.93 \\
\bottomrule
\end{tabular}
\end{table}

\subsubsection{How the penalty shape moves the price}
We now solve the loss-minimisation~\eqref{eq:objective} for the continuous design at a
representative layer $\beta=0.95$ in each region, sweeping a panel of penalties that vary the
weight $\lambda$, the curvature exponents $(p_-,p_+)$, and---in the last row---swap the power
penalty for an exponential shortfall loss $\phi_-(z)=e^{\kappa z}-1$ with linear over-pay.
Table~\ref{tab:t7} reports the calibrated trigger $d$, scale $q$, mean payout $\E[P]$, and the
realised mean basis $\E[B]$; Table~\ref{tab:t8} reports the shortfall $\E[B^+]$, over-pay
$\E[(-B)^+]$, dependence, hedging effectiveness $HE=1-\Var(L-P)/\Var(L)$, and the
shortfall capital $\VaR_{0.995}(B^+)$. A soft budget ($\mu=3$) keeps the premium anchored
unless the curvature deliberately overrides it.

\begin{figure}[H]
\centering
\includegraphics[width=0.95\textwidth]{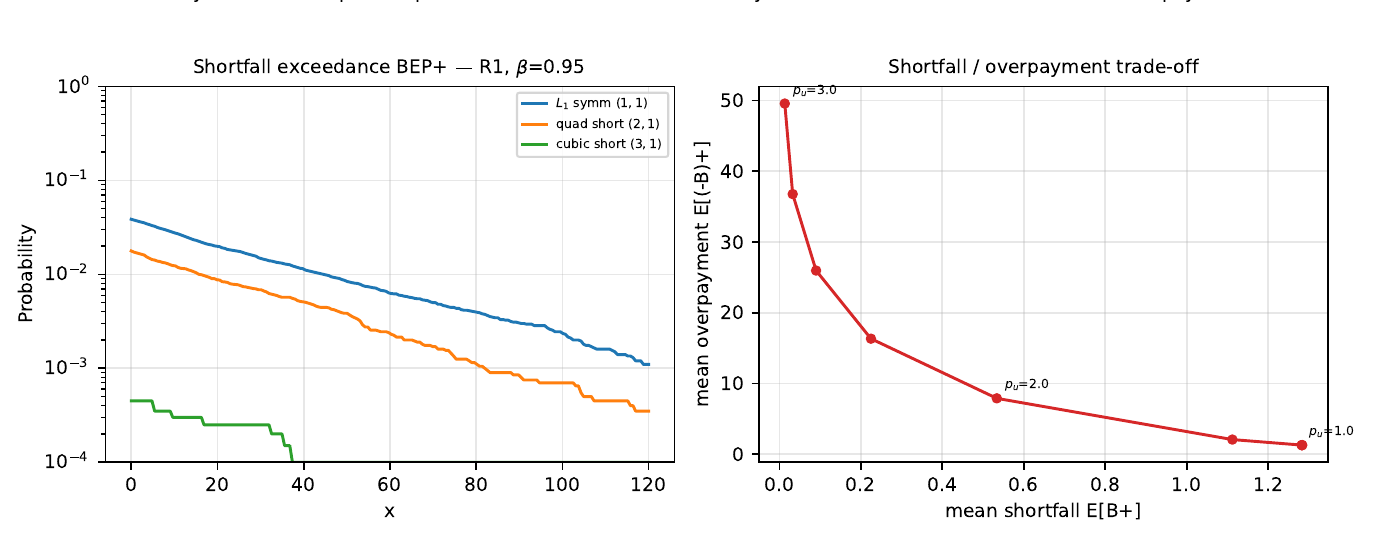}
\caption{Penalty curvature shapes the price (continuous design, R1, $\beta=0.95$). Left:
the shortfall exceedance curve $\BEP^+$ thins as the shortfall exponent rises from $1$ to $3$.
Right: the resulting shortfall/over-pay trade-off, traced as the shortfall exponent $p_-$
sweeps from $1$ to $3$ at fixed weight---heavier curvature buys a thinner shortfall tail by
accepting more over-payment.}
\label{fig:pricing}
\end{figure}

Three patterns stand out. First, \emph{weight alone barely moves the price}: the symmetric and
shortfall-tilted $L_1$ rows are nearly identical in $(d,q)$ and sit on $\E[B]=0$, because under a
linear penalty the scale direction is governed by the mean-basis condition of
Remark~\ref{rem:fair} regardless of $\lambda$. Second, \emph{curvature moves it decisively}:
raising the shortfall exponent from $1$ to $2$ to $3$ drives the scale $q$ up sharply, pushes
$\E[P]$ above the AAL, and collapses the shortfall tail---$\VaR_{0.995}(B^+)$ falls from roughly
$45$ at $(1,1)$ to near zero at $(3,1)$ in every region---while the over-pay $\E[(-B)^+]$ balloons
correspondingly. This is the shortfall/over-pay trade-off of Figure~\ref{fig:pricing}, traced by
curvature rather than by weight. Third, \emph{the general loss case behaves as expected}: the
exponential shortfall penalty produces an intermediate, region-dependent price that targets
extreme under-payment without the explosive scale of the cubic power penalty. Symmetric
curvature $(2,2)$ leaves the price essentially AAL-neutral, confirming that it is the
\emph{asymmetry} of curvature, not its presence, that relocates the premium.

\begin{table}[H]
\centering
\caption{Penalty shape and the price (continuous design, $\beta=0.95$). Columns: weight $\lambda$, curvature $(p_-,p_+)$, trigger $d$, scale $q$, mean payout $\E[P]$, mean basis $\E[B]$.}
\label{tab:t7}
\begin{tabular}{llccrrrr}
\toprule
Region & Penalty & $\lambda$ & $(p_-,p_+)$ & $d$ & $q$ & $\E[P]$ & $\E[B]$ \\
\midrule
R1 & $L_1$ symmetric & 0.50 & (1,1) & 0.591 & 117.67 & 1.77 & 0.000 \\
R1 & $L_1$ short-tilt & 0.80 & (1,1) & 0.591 & 117.85 & 1.77 & 0.000 \\
R1 & quad short & 0.80 & (2,1) & 0.435 & 243.63 & 9.14 & -7.365 \\
R1 & quad both & 0.50 & (2,2) & 0.399 & 39.30 & 1.77 & -0.000 \\
R1 & cubic short & 0.80 & (3,1) & 0.179 & 427.32 & 51.34 & -49.567 \\
R1 & exp short & 0.80 & exp & 0.709 & 432.80 & 2.68 & -0.902 \\
R2 & $L_1$ symmetric & 0.50 & (1,1) & 0.488 & 111.08 & 1.25 & -0.000 \\
R2 & $L_1$ short-tilt & 0.80 & (1,1) & 0.488 & 111.12 & 1.25 & -0.000 \\
R2 & quad short & 0.80 & (2,1) & 0.387 & 177.06 & 3.79 & -2.544 \\
R2 & quad both & 0.50 & (2,2) & 0.369 & 52.40 & 1.25 & -0.000 \\
R2 & cubic short & 0.80 & (3,1) & 0.138 & 289.01 & 23.34 & -22.094 \\
R2 & exp short & 0.80 & exp & 0.512 & 130.85 & 1.25 & 0.000 \\
R3 & $L_1$ symmetric & 0.50 & (1,1) & 0.370 & 123.35 & 1.21 & 0.000 \\
R3 & $L_1$ short-tilt & 0.80 & (1,1) & 0.370 & 123.39 & 1.21 & -0.000 \\
R3 & quad short & 0.80 & (2,1) & 0.310 & 242.48 & 3.54 & -2.329 \\
R3 & quad both & 0.50 & (2,2) & 0.294 & 74.66 & 1.21 & 0.000 \\
R3 & cubic short & 0.80 & (3,1) & 0.185 & 454.17 & 14.67 & -13.462 \\
R3 & exp short & 0.80 & exp & 0.523 & 377.22 & 1.21 & 0.000 \\
\bottomrule
\end{tabular}
\end{table}

\begin{table}[H]
\centering
\caption{Penalty shape and the basis (continuous design, $\beta=0.95$). Shortfall $\E[B^+]$, over-pay $\E[(-B)^+]$, dependence, hedging effectiveness $HE$, and shortfall capital $\VaR_{0.995}(B^+)$.}
\label{tab:t8}
\begin{tabular}{llrrrrr}
\toprule
Region & Penalty & $\E[B^+]$ & $\E[(-B)^+]$ & $\Corr$ & $HE$ & $\VaR_{0.995}(B^+)$ \\
\midrule
R1 & $L_1$ symmetric & 1.28 & 1.28 & 0.34 & 0.060 & 69.72 \\
R1 & $L_1$ short-tilt & 1.28 & 1.28 & 0.34 & 0.060 & 69.72 \\
R1 & quad short & 0.53 & 7.90 & 0.36 & -2.912 & 40.52 \\
R1 & quad both & 1.34 & 1.34 & 0.36 & 0.131 & 70.54 \\
R1 & cubic short & 0.01 & 49.58 & 0.34 & -42.709 & 0.00 \\
R1 & exp short & 1.19 & 2.09 & 0.30 & -0.731 & 68.72 \\
R2 & $L_1$ symmetric & 0.74 & 0.74 & 0.49 & 0.184 & 41.52 \\
R2 & $L_1$ short-tilt & 0.74 & 0.74 & 0.49 & 0.184 & 41.51 \\
R2 & quad short & 0.33 & 2.87 & 0.51 & -1.089 & 21.73 \\
R2 & quad both & 0.77 & 0.77 & 0.51 & 0.257 & 43.32 \\
R2 & cubic short & 0.01 & 22.11 & 0.45 & -24.590 & 0.00 \\
R2 & exp short & 0.74 & 0.74 & 0.48 & 0.143 & 40.73 \\
R3 & $L_1$ symmetric & 0.66 & 0.66 & 0.57 & 0.292 & 38.41 \\
R3 & $L_1$ short-tilt & 0.66 & 0.66 & 0.57 & 0.292 & 38.41 \\
R3 & quad short & 0.28 & 2.61 & 0.58 & -1.269 & 18.67 \\
R3 & quad both & 0.68 & 0.68 & 0.58 & 0.332 & 39.46 \\
R3 & cubic short & 0.02 & 13.48 & 0.55 & -21.318 & 0.00 \\
R3 & exp short & 0.75 & 0.75 & 0.49 & -0.308 & 40.03 \\
\bottomrule
\end{tabular}
\end{table}

\subsubsection{Portfolio risk}
The numerical assessment is based on Monte Carlo samples of regional indemnity losses
$L_r = A_r D(T_r)$. Within each region, $A_r$ and $T_r$ are independent. At the portfolio
level, regional sample pairs are combined by aligning draws across regions; unless stated
otherwise this treats the regional pairs as independent. Let $B^{\{\cdot\}}_{r,\beta} = Y_{r,\beta} - P^{\{\cdot\}}_{r,\beta}$
denote the basis in region $r$ at layer $\beta$. The aggregate and occurrence basis are
$S^{B}_\beta = \sum_{r\in\R} B^{\{\cdot\}}_{r,\beta}$ and $M^{B}_\beta = \max_{r\in\R} B^{\{\cdot\}}_{r,\beta}$.

\begin{figure}[H]
\centering
\includegraphics[width=0.92\textwidth]{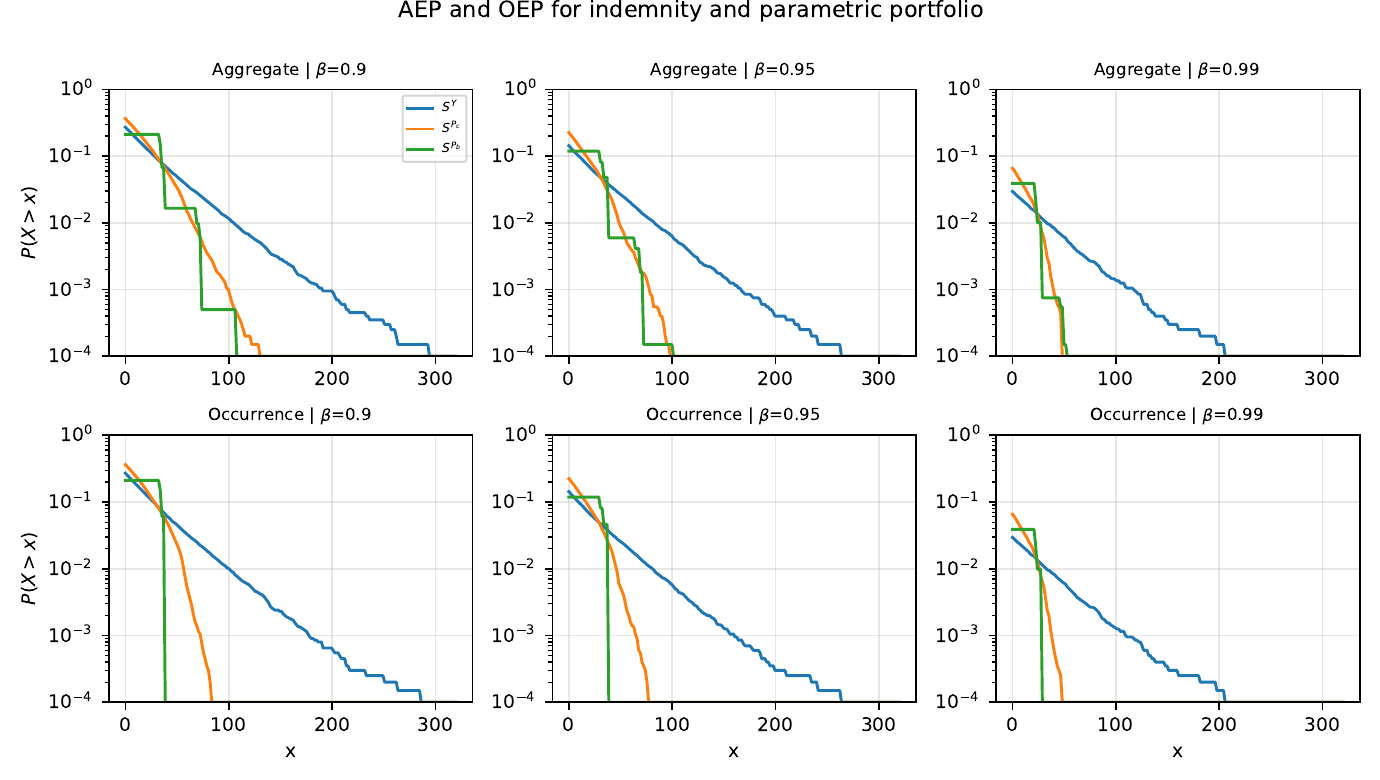}
\caption{AEP and OEP for the indemnity and parametric portfolio.}
\label{fig:port_ep}
\end{figure}

\begin{figure}[H]
\centering
\includegraphics[width=0.92\textwidth]{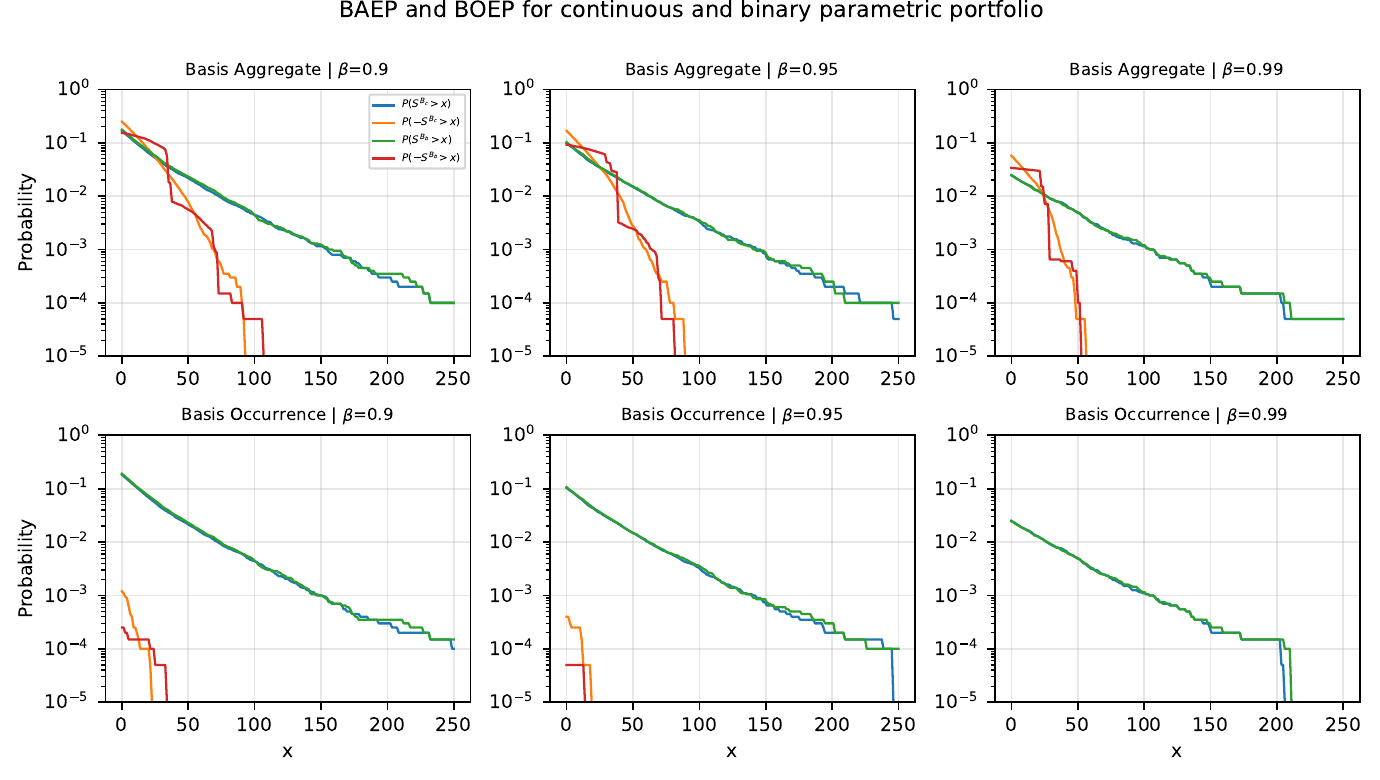}
\caption{BAEP and BOEP for the continuous and binary parametric portfolio.}
\label{fig:port_basis}
\end{figure}

\begin{figure}[H]
\centering
\includegraphics[width=0.92\textwidth]{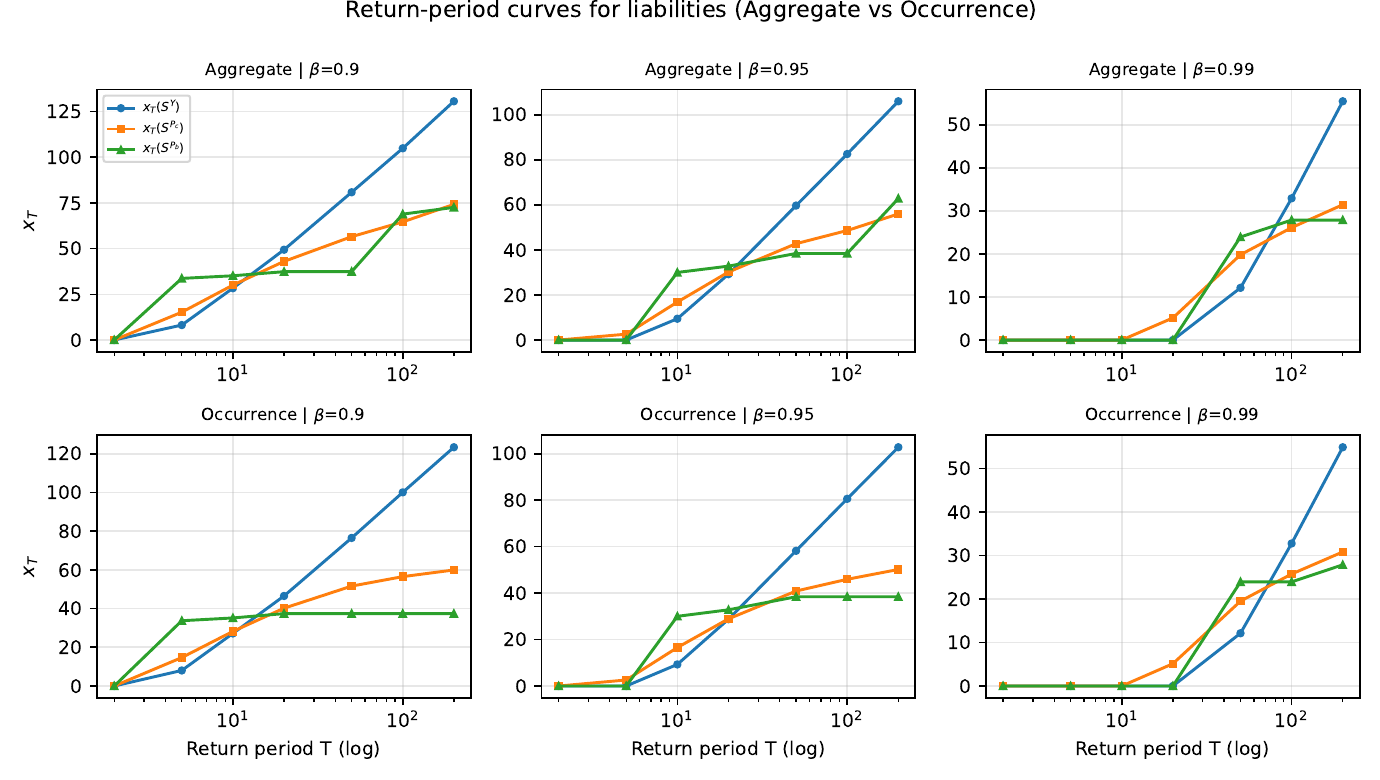}
\caption{Return-period curves for liabilities (Aggregate vs Occurrence).}
\label{fig:port_rp_liab}
\end{figure}

\begin{figure}[H]
\centering
\includegraphics[width=0.92\textwidth]{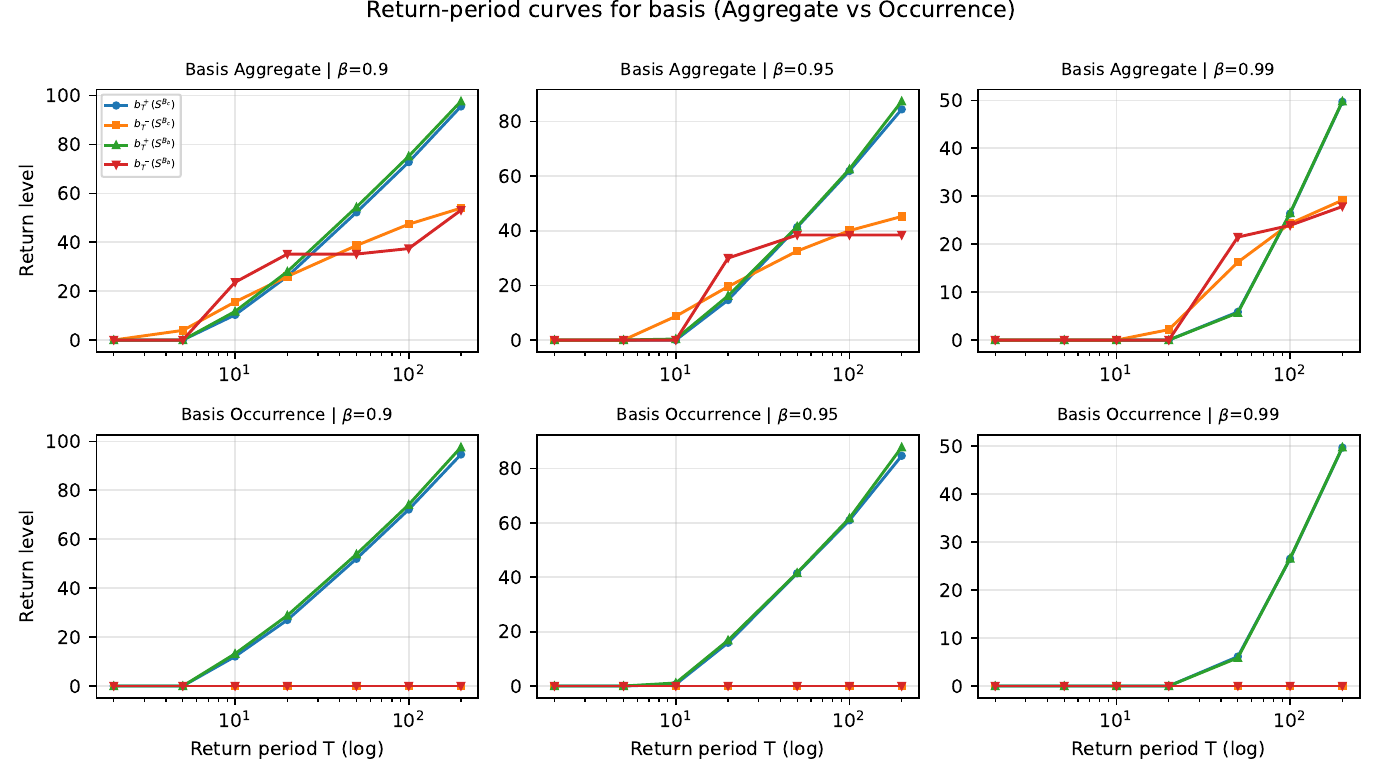}
\caption{Return-period curves for basis (Aggregate vs Occurrence, continuous \& binary, $\pm$).}
\label{fig:port_rp_basis}
\end{figure}

Tables~\ref{tab:t4} and~\ref{tab:t5} report the aggregate-vs-occurrence level/dependence and
point diagnosis. Table~\ref{tab:t6} reports portfolio capital metrics. Figures~\ref{fig:port_basis},
\ref{fig:port_rp_liab}, and~\ref{fig:port_rp_basis} report the basis exceedance and
return-period curves.

\begin{table}[H]
\centering
\caption{Aggregate vs Occurrence: level and dependence.}
\label{tab:t4}
\begin{tabular}{llcrrr}
\toprule
Scope & Design & $\beta$ & $\E[Y]$ & $\Corr(Y,P)$ & $\Var(B)$ \\
\midrule
Agg & b & 0.90 & 8.05 & 0.49 & 386.64 \\
Agg & b & 0.95 & 4.23 & 0.40 & 250.97 \\
Agg & b & 0.99 & 0.94 & 0.22 & 70.18 \\
Agg & c & 0.90 & 8.05 & 0.55 & 338.01 \\
Agg & c & 0.95 & 4.23 & 0.45 & 224.02 \\
Agg & c & 0.99 & 0.94 & 0.24 & 67.12 \\
Occ & b & 0.90 & 7.68 & 0.47 & 221.56 \\
Occ & b & 0.95 & 4.12 & 0.38 & 152.09 \\
Occ & b & 0.99 & 0.93 & 0.22 & 49.76 \\
Occ & c & 0.90 & 7.68 & 0.53 & 207.35 \\
Occ & c & 0.95 & 4.12 & 0.44 & 147.02 \\
Occ & c & 0.99 & 0.93 & 0.23 & 48.81 \\
\bottomrule
\end{tabular}
\end{table}

\begin{table}[H]
\centering
\caption{Aggregate vs Occurrence: point diagnosis.}
\label{tab:t5}
\begin{tabular}{llcrrrrr}
\toprule
Scope & Design & $\beta$ & $\PR(P>0)$ & $\PR(B>0)$ & $\PR(B<0)$ & $\E[B^+]$ & $\E[(-B)^+]$ \\
\midrule
Agg & b & 0.90 & 0.21 & 0.18 & 0.15 & 4.24 & 4.24 \\
Agg & b & 0.95 & 0.12 & 0.10 & 0.09 & 2.64 & 2.64 \\
Agg & b & 0.99 & 0.04 & 0.02 & 0.03 & 0.76 & 0.76 \\
Agg & c & 0.90 & 0.36 & 0.17 & 0.25 & 3.95 & 3.95 \\
Agg & c & 0.95 & 0.22 & 0.10 & 0.17 & 2.55 & 2.55 \\
Agg & c & 0.99 & 0.07 & 0.02 & 0.06 & 0.76 & 0.76 \\
Occ & b & 0.90 & 0.21 & 0.19 & 0.00 & 4.40 & 0.00 \\
Occ & b & 0.95 & 0.12 & 0.11 & 0.00 & 2.70 & 0.00 \\
Occ & b & 0.99 & 0.04 & 0.03 & 0.00 & 0.77 & 0.00 \\
Occ & c & 0.90 & 0.36 & 0.18 & 0.00 & 4.14 & 0.01 \\
Occ & c & 0.95 & 0.22 & 0.10 & 0.00 & 2.63 & 0.00 \\
Occ & c & 0.99 & 0.07 & 0.02 & 0.00 & 0.77 & 0.00 \\
\bottomrule
\end{tabular}
\end{table}

\begin{table}[H]
\centering
\caption{Capital metrics for portfolio of NatCat layer $Y_{r,\beta}$ and NatPar liabilities $P^{\{\cdot\}}_{r,\beta}$: $\VaR_{0.995}$ and $\TVaR_{0.995}$, by layer $\beta$.}
\label{tab:t6}
\begin{tabular}{llrrrrrr}
\toprule
Scope & $\beta$ & $\VaR(Y_\beta)$ & $\TVaR(Y_\beta)$ & $\VaR(P^c_\beta)$ & $\TVaR(P^c_\beta)$ & $\VaR(P^b_\beta)$ & $\TVaR(P^b_\beta)$ \\
\midrule
Agg & 0.90 & 130.65 & 169.28 & 74.23 & 88.68 & 72.59 & 75.50 \\
Agg & 0.95 & 105.98 & 145.01 & 55.97 & 69.71 & 62.81 & 68.29 \\
Agg & 0.99 & 55.43 & 91.02 & 31.42 & 36.54 & 27.86 & 29.42 \\
Occ & 0.90 & 123.37 & 158.98 & 59.96 & 66.77 & 37.42 & 37.42 \\
Occ & 0.95 & 102.80 & 138.95 & 50.17 & 58.49 & 38.43 & 38.43 \\
Occ & 0.99 & 54.86 & 89.54 & 30.81 & 35.20 & 27.86 & 27.86 \\
\bottomrule
\end{tabular}
\end{table}

\section{Results interpretation of the frost case study}\label{sec:interp}
Figures~\ref{fig:ep_indiv}--\ref{fig:basis_indiv} reveal a systematic pattern: in the far tail, the shortfall component of basis
risk dominates the overpayment component. Interpreting this correctly requires separating
an economic conclusion (who benefits) from a mathematical one (why the tail geometry
looks the way it does). The plots primarily reflect a structural feature of the design: the
indemnity-layer benchmark $Y_{r,\beta}$ is driven by exposure--severity through $L = AD$, while the
parametric payouts are functions of the damage index $D$ alone and are therefore bounded.

\paragraph{The core mechanism.} In our setting $L = A_r D_r$, where the random exposure/severity
multiplier $A_r > 0$ is independent of $D_r$ and may exhibit substantial right-tail variability,
while the damage ratio satisfies $D_r \in [0, 1]$. The NatPar designs are functions of $D$ and
inherit its bounded support: $0 \le P^b_{r,\beta} \le q^b_{r,\beta}$ and $0 \le P^c_{r,\beta} \le q^c_{r,\beta}(1 - d^c_{r,\beta})$,
where $d^{\{\cdot\}}_{r,\beta}$ is the optimal trigger and $q^{\{\cdot\}}_{r,\beta}$ the AAL-neutral scale that matches the mean layer loss $\E[Y_{r,\beta}]$. In
contrast, $Y_{r,\beta} = (A_r D_r - l_{r,\beta})^+$ can be large whenever $A_r$ is large enough, even if $D_r$ is
only moderate. Consequently, in extreme states the basis behaves as $B^{\{\cdot\}}_{r,\beta} \approx Y_{r,\beta}$ whenever
$Y_{r,\beta} \gg \sup P^{\{\cdot\}}_{r,\beta}$, so the far-right tail of $B^{\{\cdot\}}_{r,\beta}$ inherits the tail of $Y_\beta$ while the overpayment side is capped.
This is the mathematical reason the return-period shortfall curves grow faster than the
corresponding overpayment curves in Figure~\ref{fig:rp_basis}, and why the basis exceedance
curves $\BEP^+$ remain comparatively heavy in Figure~\ref{fig:basis_indiv}.

\paragraph{Unspanned severity.} The dominance of shortfall is not merely ``because $P$ is bounded'';
it arises because a $D$-only payout cannot span exposure-driven severity. Large values of
$Y_{r,\beta}$ can occur via two mechanisms: (i) $A_r$ large with moderate $D_r$, or (ii) $D_r$ large with
moderate $A_r$. Binary and continuous NatPar designs respond only to mechanism (ii). Under
independence of $A$ and $D$, mechanism (i) occurs with non-negligible probability in the tail,
producing states where $Y_{r,\beta} > 0$ but $P^{\{\cdot\}}_{r,\beta}$ is small or zero, hence $B^{\{\cdot\}}_{r,\beta}$ is large and positive
(shortfall). Overpayment requires the reverse mismatch: $D$ large together with small $A_r$ so
that $Y_{r,\beta}$ is near zero while $P^{\{\cdot\}}_{r,\beta}$ is positive. Such states exist, but their severity is capped by
$\sup P^{\{\cdot\}}_{r,\beta}$, so overpayment does not dominate at long return periods.

\paragraph{Profitability.} It is tempting to read ``shortfall dominates overpayment'' as an issuer
advantage and therefore a reason to underwrite the product. That inference is incomplete.
Overpayment corresponds to paying more than the benchmark layer and is the issuer's
direct economic downside relative to $Y_{r,\beta}$. Shortfall, however, is not a cash loss to the
issuer; it is a coverage failure borne by the policyholder and often a commercial or conduct
risk for the issuer. A design that generates large tail shortfalls may be capital-tractable
(bounded liability) but can be commercially fragile. Hence tail shortfall dominance is
primarily evidence of bounded-liability structure rather than a direct profitability statement.

\paragraph{Design implication.} We select $q^{\{\cdot\}}_{r,\beta}$ to match $\E[P^{\{\cdot\}}_{r,\beta}] = \E[Y_{r,\beta}]$ (AAL matching at the
layer). Figures~\ref{fig:ep_indiv} and~\ref{fig:rp_liab} show that this mean matching can coexist with large differences
in tail behaviour, because matching $\E[Y_{r,\beta}]$ does not control how $Y_\beta$ decomposes into
exposure-driven versus damage-driven extremes. When exposure variability dominates
severity, a damage-only index necessarily under-replicates the far tail.

\paragraph{Individual risk: what the three regions show.} The individual capital table
(Table~\ref{tab:t3}) shows that the three regional assets behave differently in the far tail and
that the difference is design-dependent. For the most frost-exposed region $\text{R1}$ at
$\beta=0.90$, the benchmark layer remains heavy-tailed with $\VaR_{0.995}(Y_\beta)\approx 106$ and
$\TVaR_{0.995}(Y_\beta)\approx 145$, while the optimally-triggered continuous design tracks the
body but caps the tail ($\VaR_{0.995}(P^c_\beta)\approx 55$, $\TVaR_{0.995}(P^c_\beta)\approx 57$)
and the binary design is more strongly capped ($\VaR_{0.995}(P^b_\beta)\approx 35$). The graded
severity across regions---$\text{R1}$ heaviest, then $\text{R3}$ and $\text{R2}$---is produced
entirely by the hazard law, since the damage curve is identical. At $\beta=0.99$ the loss layer is
thin (e.g.\ $\VaR_{0.995}(Y_\beta)\approx 18$ for $\text{R2}$), and the digital binary payout,
sized to be AAL-neutral against that thin layer, can carry a liability \emph{as large as} the
indemnity layer's tail capital and larger than the continuous design's
($\VaR_{0.995}(P^b_\beta)\approx 28$ for $\text{R2}$, against $\VaR_{0.995}(Y_\beta)\approx 18$ and
$\VaR_{0.995}(P^c_\beta)\approx 21$): the point mass lands near the layer's $99.5$th percentile.
This is the capital cost of digitisation made explicit, and it is a feature of payout shape, not of
mean level. The optimal trigger keeps all three regions responsive at high $\beta$---the continuous
design tracks the layer closely ($\VaR_{0.995}(P^c_\beta)\approx 21$--$25$ against
$\VaR_{0.995}(Y_\beta)\approx 18$--$24$ across regions)---so optimisation avoids the tail
non-spanning that an arbitrary high-percentile trigger would have produced. Because the damage
curve is identical across regions, these differences isolate the effect of the hazard law on the
tail.

\paragraph{Portfolio versus individual risk.} Once we move to the three-asset portfolio, the
point-diagnosis tables clarify what pooling does and does not fix. First, the mean layer size
drops sharply as $\beta$ increases ($\E[S^Y]\approx 9.0$ at $\beta=0.90$, $4.9$ at $\beta=0.95$, and
$1.2$ at $\beta=0.99$); this is the mechanical effect of pushing the attachment deeper into the
tail, not a diversification effect. The portfolio story is instead in dependence and mismatch.
The continuous design remains systematically more aligned with the benchmark than the
binary design: in the aggregate (AEP-style) view, $\Corr(S^Y, S^P)$ is about $0.71$ for design
$c$ versus $0.62$ for design $b$ at $\beta=0.90$, narrowing toward $0.33$/$0.32$ as
$\beta\to 0.99$ where both designs become sparse. More importantly for governance, the
portfolio basis dispersion is higher for the binary design: $\Var(S^B)$ is about $397$ ($b$)
versus $306$ ($c$) at $\beta=0.90$, and the occurrence dispersion $\Var(M^B)$ is about $238$
($b$) versus $198$ ($c$). This is the numerical counterpart of what the basis-tail plots
suggest: the binary contract is not just capped, it is capped in a way that concentrates the
unspanned component into the shortfall side whenever exposure-driven severity dominates.

The same conclusion holds under the occurrence (OEP-style) lens. The occurrence view is the
one that bites for suitability and operational stress: even if the aggregate mismatch nets out
across assets in some years, a single region can still experience a large shortfall. That is why,
in addition to reporting the aggregate basis $S^B = \sum_r B_r$, we recommend an occurrence
basis $M^B := \max_r B_r$ as a regulatory object. Importantly, this definition should be based on
the per-asset basis terms (not portfolio-level quantiles of $L$ and $D$), because attachment and
triggers are contractual at the risk level; computing ``portfolio'' $l^{\text{port}}_\beta$ and $d^{\text{port}}_\beta$
mixes attachment/triggers across assets and can mask the very cross-asset tail asymmetries the
portfolio analysis is meant to reveal. Overall, the portfolio does not reverse the individual
diagnosis; it sharpens it.

\paragraph{Pricing implication.} Tables~\ref{tab:t7}--\ref{tab:t8} add the pricing lens to this picture.
The premium is governed far more by the \emph{curvature} of the mismatch penalty than by its
weight: shortfall-tilting the weight under a linear penalty leaves the price on the AAL-neutral
value, whereas convex shortfall curvature lifts the scale and thins the shortfall tail at the cost
of overpayment, tracing the trade-off in Figure~\ref{fig:pricing}. The AAL-neutral (``fair'')
price is recovered only at the symmetric-linear corner; it is a useful reference point, not a
distinct objective. For binary triggers the single trigger is pinned by the budget alone, so
curvature has no purchase---a useful guardrail against over-claiming pricing flexibility for
digital designs.

\paragraph{Regulatory and reporting implication.} The basis diagnostics provide information that
cannot be inferred from AAL and standard EP curves alone. Basis exceedance curves
$\BEP^\pm$ summarise the frequency of extreme shortfall/overpayment events as a function of
severity, while basis return levels $b^\pm_T$ translate those exceedance probabilities into $T$-year
basis severities. These make explicit whether a parametric design is best interpreted as a loss
proxy with acceptable tail mismatch, or as a bounded liquidity instrument whose tail
protection is limited by construction. We therefore view the pair $(\BEP^\pm, b^\pm_T)$ as essential
complements to the usual NatCat objects.

\paragraph{Market benefit: liquidity.} A practical benefit of the bounded, index-linked structure
is not that it perfectly replicates tail loss, but that it can function as a contingent liquidity
instrument in markets where high exposure uncertainty creates insurability gaps. When
exposure variability is large or hard to verify quickly, indemnity-style settlement becomes
slow and costly. A parametric contract tied to an observable hazard/damage index $D$ can
still be offered because the liability is auditable and capped. The relevant question is not
whether $\BEP^+$ can be made negligible, but whether the trigger and scale deliver timely cash
in the states where liquidity constraints bind, and whether residual loss risk is transparently
disclosed and, where needed, complemented by additional layers.

\section{Who gains, and when: a temporal reading of basis risk}\label{sec:whogains}
The basis return-period curves of Section~\ref{sec:frost} support an economic reading that, to
our knowledge, is not made explicit in the design-first literature: under an AAL-neutral
parametric contract, insurer and insuree are not made indifferent but are placed on
\emph{opposite sides of the time axis}. This section develops that reading and connects it to
why NatPar has grown in popularity as catastrophe risk has intensified.

\subsection{The crossover horizon}
\begin{figure}[ht]
\centering
\includegraphics[width=0.95\textwidth]{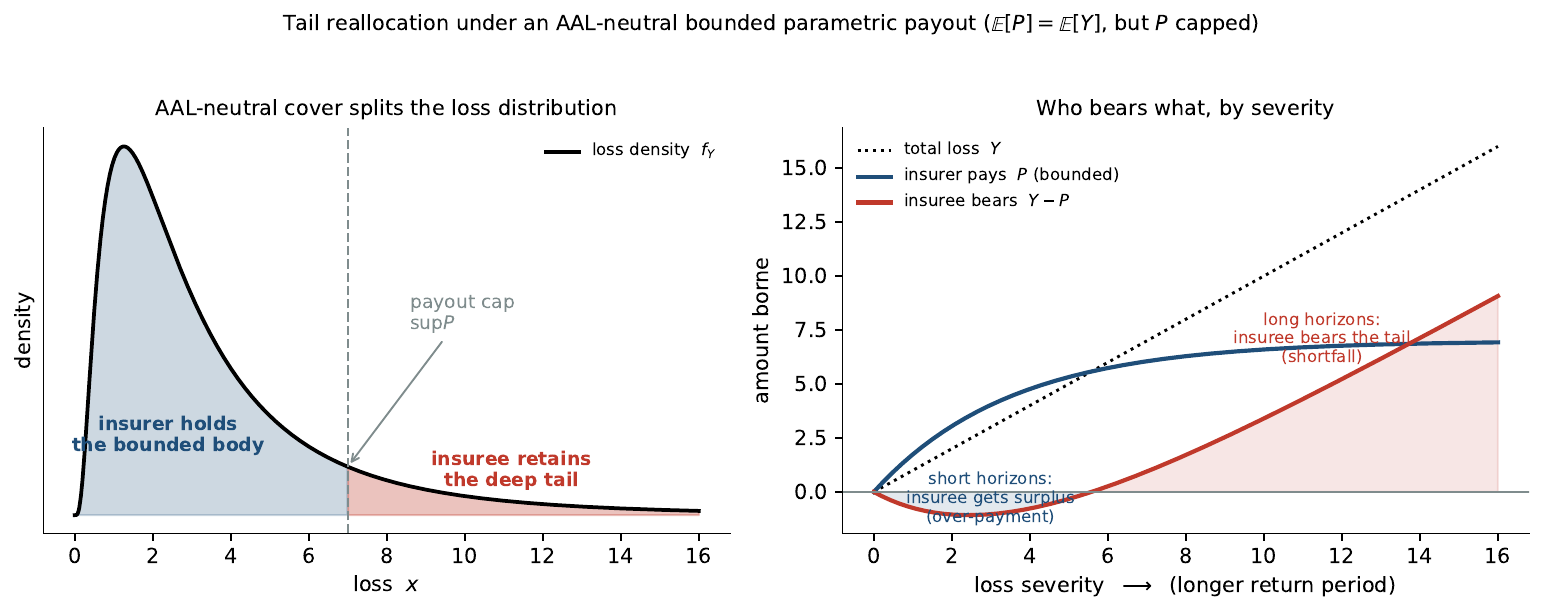}
\caption{How an AAL-neutral bounded parametric payout reallocates the loss between insurer and
insuree (schematic). Left: the payout caps at $\sup P$, so the insurer absorbs the body of the
loss density while the insuree retains the deep tail beyond the cap; the means are equalised but
the tail is not. Right: as severity (return period) grows, the parametric payout $P$ saturates at
its cap while the indemnity loss $Y$ keeps rising, so the insuree's retained loss $Y-P$ starts
\emph{negative} (a short-horizon surplus from over-payment) and then crosses zero and climbs (a
long-horizon shortfall). The crossover is the horizon at which the advantage flips.}
\label{fig:realloc}
\end{figure}

Fix the basis $B=Y-P$, so $B>0$ is shortfall (the insuree is under-covered) and $B<0$ is
over-payment (the insurer pays more than the realised loss). Under the symmetric-linear
optimal trigger the mean basis is zero and the two average severities coincide
($\E[B^+]=\E[B^-]$), so over a full statistical lifetime neither party has an edge---this is
exactly the content of Remark~\ref{rem:relocate}. The asymmetry is entirely in \emph{timing},
and the basis return-period curves (Figure~\ref{fig:rp_basis}) make it precise.
Figure~\ref{fig:realloc} gives the schematic intuition before the empirical curves: a bounded,
AAL-neutral payout covers the body of the loss distribution but cannot follow the deep tail, so
the insurer holds the bounded middle while the insuree retains the extreme, and---reading by
severity---the insuree moves from a short-horizon surplus to a long-horizon shortfall.

At short and moderate return periods the over-payment level turns positive before the
shortfall level does. For the continuous design at $\text{R1}$, $\beta=0.95$, the over-payment
return level is already material at a $25$-year horizon ($b^-_{25}\approx 13$) while the
shortfall is still building, becoming material only at longer horizons. The mechanism is
structural: over-payment occurs in the \emph{common} states where the index trips and pays a
bounded amount but the realised loss is small, whereas shortfall requires the \emph{rarer}
states in which exposure-driven severity outruns the capped payout. In the horizon band a
policyholder actually experiences over a normal programme---the lower tens of years---the
parametric contract therefore tends to pay them \emph{more} than their realised loss. The
insuree benefits in the short run.

Beyond a crossover horizon the ordering reverses and shortfall dominates. Across the three
regions the continuous-design crossover sits in the range $T\approx 60$--$90$ years
($\text{R1}\approx 61$, $\text{R2}\approx 89$, $\text{R3}\approx 91$); past it, the shortfall
return level exceeds the over-payment level and keeps climbing, reaching $b^+_{200}\approx 70$ at
$\text{R1}$ against an over-payment level of $\approx 38$. The binary design crosses earlier and
far more sharply, because its over-payment is frozen at the fixed payout while its shortfall
continues to grow: a digital payout simply cannot scale with a once-in-two-century exposure shock.
In the deep tail it is the \emph{insuree} who is exposed---the bounded payout falls progressively
further behind the unbounded indemnity loss, so the rare catastrophic year is borne by the
policyholder, not the issuer. The crossover return period---several decades for the continuous
design here, and materially shorter for the binary design---is the natural single number to report
as the horizon at which the advantage flips.

\subsection{Why this grows with catastrophe intensity}
This temporal split is the key to a motivation for NatPar that is often asserted but rarely made
precise. As catastrophe risk has intensified---larger exposures from economic growth, and
more frequent and more severe hazard states from climate change---the right tail of $Y=A\,D$
has fattened: growth thickens the tail of exposure $A$, and climate change shifts and thickens
the tail of damage $D$. A bounded parametric payout does not follow that tail. Holding the
contract AAL-neutral, the consequence is not a change in the average transfer but an
\emph{enlargement of the deep-tail region that lies past the crossover horizon}---precisely the
region the insurer sheds and the insuree absorbs. The heavier the catastrophe tail becomes,
the more severe the post-crossover shortfall, and so the more the bounded structure relocates
toward the insuree. This gives a clean, model-based statement of the popular intuition:
\emph{as catastrophe risk grows, an AAL-neutral parametric contract increasingly hands the
insurer the manageable middle of the loss distribution while leaving the increasingly heavy tail
with the insuree.}

It is important, however, to separate three distinct effects that the bounded structure delivers
simultaneously, because only one of them is a transfer to the insuree, and conflating them
overstates the case.
\begin{enumerate}[leftmargin=1.6em]
\item \textbf{Tail transfer (a cost to the insuree).} Deep-tail severity past the crossover
horizon moves from insurer to policyholder, as above. This grows with catastrophe intensity
and is the channel that raises suitability and conduct concerns.
\item \textbf{Capital and uncertainty relief (not a transfer).} Replacing exposure-driven
indemnity loss with an index-driven bounded payout removes exposure uncertainty and
development risk from the insurer's liability, lowering VaR/TVaR per unit of expected loss and
shortening the reserving tail. This relief comes from boundedness, verifiability, and the
absence of a loss-adjustment tail; it would exist even if basis risk were zero, and it is the
channel that makes the product acceptable to capital providers and supervisors.
\item \textbf{Liquidity (a benefit to the insuree).} Fast, auditable, index-verified settlement
delivers timely cash in exactly the states where liquidity constraints bind---a genuine benefit
to the policyholder that an indemnity contract, with slow loss adjustment, does not provide.
\end{enumerate}
A complete account of NatPar's rise must credit all three. The capital and liquidity channels
explain why the product clears at all; the tail-transfer channel explains why its popularity
tracks catastrophe intensity and why it warrants governance attention. Read this way, NatPar's
growth and the widening climate protection gap are two faces of the same phenomenon: when
the loss tail grows faster than capacity and model certainty can follow, the market clears by
bounding what is contractual---retreating to the part of the distribution that remains insurable
under binding capital and uncertainty constraints, and leaving the rest shifted or uninsured.

\subsection{Diversification and the relocation of the tail}\label{sec:diversification}
\begin{figure}[H]
\centering
\includegraphics[width=0.95\textwidth]{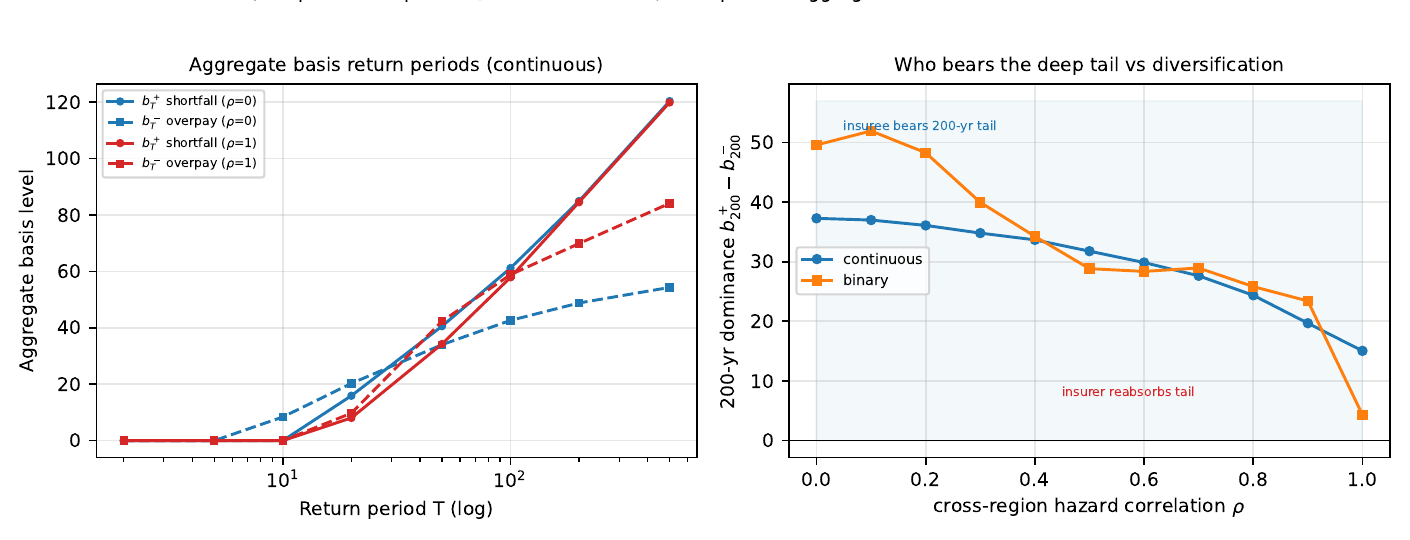}
\caption{Diversification and the relocation of the tail (independent exposures, correlated
hazard). Left: aggregate-basis return-period curves for independent ($\rho=0$, blue) versus
comonotonic ($\rho=1$, red) hazard; under independence the shortfall side (solid) dominates the
tail, while under comonotonicity the overpayment side (dashed) catches up and overtakes it.
Right: the $200$-year dominance margin $b^+_{200}-b^-_{200}$ as a function of $\rho$; positive
values mean the insuree bears the deep tail, negative values mean the insurer reabsorbs it. The
binary design is the more dependence-sensitive.}
\label{fig:diversification}
\end{figure}

The crossover horizon of Section~\ref{sec:whogains} was computed risk-by-risk. A portfolio
adds a second force---diversification---that interacts with the relocation in a way that is itself
informative for both pricing and supervision. We isolate it with a controlled experiment.
Exposures are taken \emph{independent across regions} (the dollar value at risk in one region
says nothing about another), while the hazard carries the only cross-region dependence through
a one-factor model: each regional index is $Z_r=\sqrt{\rho}\,F+\sqrt{1-\rho}\,\varepsilon_r$ with a
common climate factor $F$ and idiosyncratic $\varepsilon_r$, so $\rho$ tunes the probability that
cold extremes strike several regions in the same season. Each design is re-priced at the
AAL-neutral optimal trigger within each region, and we read the \emph{aggregate} basis
$S^B=\sum_r(Y_r-P_r)$. Table~\ref{tab:t9} and Figure~\ref{fig:diversification} report the result.

\paragraph{Diversification smooths the body but not the tail, pulling the crossover earlier.}
Under independent hazard ($\rho=0$) the aggregate-basis crossover for the continuous design
falls to roughly $30$--$40$ years, shorter than the $\approx 60$--$90$-year standalone crossover
of a single region. The reason is asymmetric. The frequent, mild over-payments are largely
idiosyncratic across regions and partly net out when summed, so the \emph{body} of the
aggregate basis is compressed by pooling. The severe shortfalls, by contrast, sit in the heavy
$A\,D$ tail and do not cancel: a bad year in one region still contributes its full unspanned loss
to the sum. Diversification therefore thins the over-payment side faster than the shortfall side,
and the shortfall overtakes over-payment sooner in aggregate than it does for a lone risk. For
the insuree this is a double-edged outcome: pooling stabilises the routine over-payment surplus
they enjoyed at short horizons, but it also brings forward the horizon at which they begin to
bear net shortfall.

\paragraph{Correlated hazard erodes the diversification and hands the deep tail back to the
insurer.} As $\rho$ rises the common climate factor synchronises regional cold extremes, so
the regions stop diversifying in precisely the states that matter. The right panel of
Figure~\ref{fig:diversification} tracks the $200$-year dominance margin
$b^+_{200}-b^-_{200}$: it is firmly positive at low $\rho$ (the insuree bears the deep tail, as in
the single-risk analysis) but falls toward zero and turns \emph{negative} as $\rho\to1$. Under
near-comonotonic hazard the bounded payouts of all regions trigger together, stacking into a
large aggregate payout in the common catastrophic year while the realised aggregate loss, though
large, is matched or exceeded by that stacked payout; the insurer then \emph{over-pays} in the
tail and reabsorbs the very cost that independence had relocated to the insuree. Table~\ref{tab:t9}
shows the aggregate basis variance, which is essentially the sum of standalone variances at
$\rho=0$ (the diversification ratio is $\approx 1.00$, confirming near-independence in the base
calibration), rising above it as $\rho\to1$ (ratio $\approx 1.08$ continuous, $\approx 1.26$ binary):
correlation destroys the pooling benefit and concentrates risk back into the aggregate tail.

\paragraph{Implications.} Three points follow. First, the tail relocation of
Section~\ref{sec:whogains} is not a fixed property of a contract but depends on the portfolio it
sits in and on the dependence among hazards: the same AAL-neutral design can leave the
deep tail with the insuree on a diversified, weakly-correlated book and with the insurer on a
concentrated, highly-correlated one. Second, the binary design is the more dependence-sensitive
of the two---its dominance margin swings furthest as $\rho$ moves---because its fixed payouts
stack most mechanically under comonotonicity; digitisation that looks capital-light on an
independent book can become capital-heavy on a correlated one. Third, and most relevant under
climate change, the spatial correlation of extremes is itself increasing: large-scale climate
drivers raise $\rho$ over time, which by this mechanism partially \emph{returns} deep-tail cost to
insurers even as growing exposure fattens the tail in the first place. The two climate forces
therefore push in opposite directions on who ultimately bears the tail, and the occurrence-basis
object $M^B$ remains the right supervisory lens because it captures the worst single-region
shortfall regardless of how the aggregate nets out.

\begin{table}[H]
\centering
\caption{Diversification and the aggregate basis (continuous and binary, $\beta=0.95$, independent exposures). Crossover return period; $200$- and $50$-year shortfall ($b^+$) and overpayment ($b^-$) levels; and the diversification ratio $\Var(S^B)/\sum_r\Var(B_r)$, as the cross-region hazard correlation $\rho$ varies.}
\label{tab:t9}
\begin{tabular}{lccrrrrr}
\toprule
$\rho$ & Design & crossover $T$ & $b^+_{50}$ & $b^-_{50}$ & $b^+_{200}$ & $b^-_{200}$ & div.\ ratio \\
\midrule
0.0 & c & 32 & 40.56 & 34.01 & 84.92 & 48.80 & 1.00 \\
0.0 & b & 10 & 42.23 & 32.37 & 88.58 & 34.77 & 1.00 \\
0.3 & c & 34 & 39.03 & 33.69 & 83.77 & 49.71 & 1.00 \\
0.3 & b & 38 & 40.40 & 32.15 & 86.05 & 45.82 & 0.99 \\
0.6 & c & 37 & 37.63 & 33.57 & 83.21 & 53.36 & 1.00 \\
0.6 & b & 12 & 39.22 & 34.19 & 85.16 & 62.21 & 0.99 \\
1.0 & c & 105 & 34.25 & 42.32 & 84.53 & 69.88 & 1.05 \\
1.0 & b & 15 & 35.22 & 44.85 & 84.79 & 76.53 & 1.05 \\
\bottomrule
\end{tabular}
\end{table}

\subsection{Tail dependence, not correlation, governs the deep-tail relocation}\label{sec:taildep}
The diversification experiment of Section~\ref{sec:diversification} used a one-factor Gaussian
model, in which cross-region dependence is summarised by a single linear correlation $\rho$. For
catastrophe perils this understates the danger in exactly the region that matters. Correlation is
a whole-distribution, linear measure, and the Gaussian copula has \emph{zero tail dependence}:
the conditional probability that two regions hit their extremes together vanishes as one moves
further into the tail, however high $\rho$ is set. Yet the defining feature of climate perils is the
opposite---a continental cold outbreak drives regions to their cold extremes \emph{jointly}. The
relevant question is therefore not how correlated the regions are on average, but how strongly
they fail together in the tail.

\begin{figure}[H]
\centering
\includegraphics[width=0.95\textwidth]{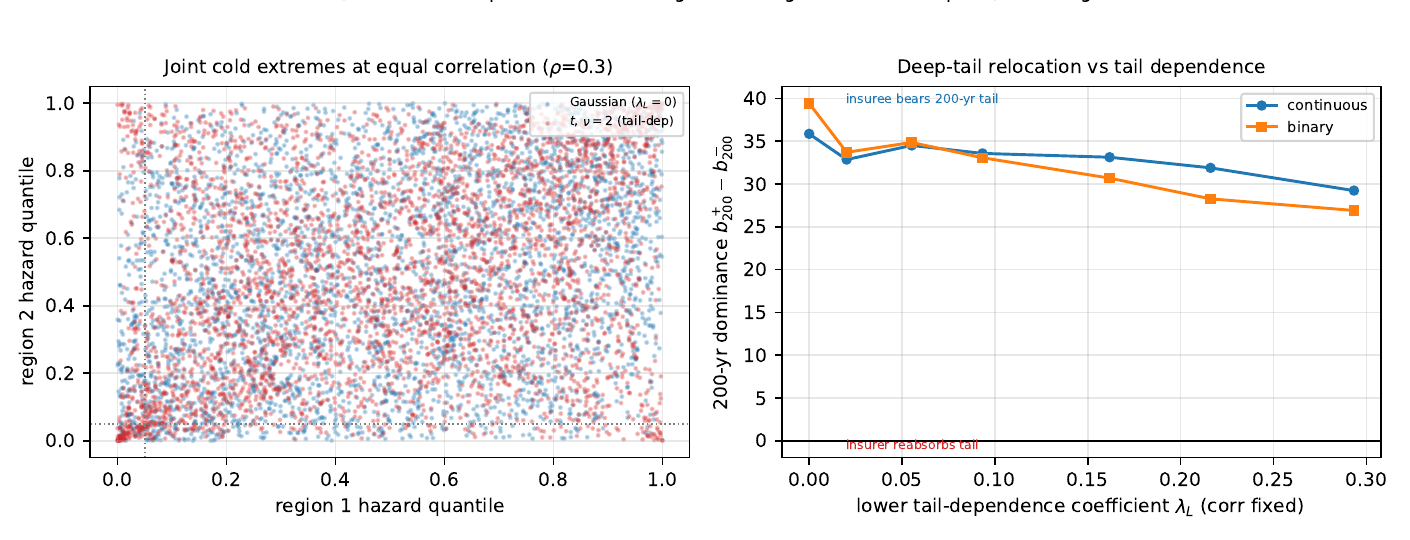}
\caption{Tail dependence, not correlation, governs the deep-tail relocation. Left: regional
hazard-quantile draws under a Gaussian copula ($\lambda_L=0$) and a $t$-copula with $\nu=2$,
both at the same linear correlation $\rho\approx0.3$; the $t$-copula concentrates mass in the
joint lower-left corner (joint cold extremes) that the Gaussian misses. Right: the $200$-year
dominance margin $b^+_{200}-b^-_{200}$ as a function of the lower tail-dependence coefficient
$\lambda_L$ at fixed correlation; heavier tail dependence monotonically shifts the deep tail from
insuree toward insurer, with the binary design the more sensitive.}
\label{fig:taildep}
\end{figure}

We isolate this by holding the linear correlation fixed and varying only the coefficient of lower
tail dependence $\lambda_L$, replacing the Gaussian copula on the regional hazards with a
$t$-copula whose degrees of freedom $\nu$ control $\lambda_L$ (smaller $\nu$, heavier tail
dependence; the Gaussian is the $\nu\to\infty$ limit with $\lambda_L=0$). As noted in
Remark~\ref{rem:catnature}, this changes only the copula: each regional temperature stays exactly
Gaussian, so tail dependence is introduced without giving any single hazard a heavy marginal tail.
Exposures remain
independent across regions, and frost is mapped as a lower extreme so that the copula's joint
tail corresponds to joint cold. Table~\ref{tab:t10} and Figure~\ref{fig:taildep} report the effect.

\paragraph{Correlation can be blind to the joint catastrophe.} The left panel of
Figure~\ref{fig:taildep} shows draws of the regional hazard quantiles under a Gaussian copula
and under a $t$-copula with $\nu=2$, calibrated to the \emph{same} linear correlation
$\rho\approx0.3$. The two are nearly indistinguishable through the body, but the $t$-copula packs
visibly more mass into the joint lower-left corner---both regions simultaneously in their cold
extremes. Table~\ref{tab:t10} quantifies this: as $\lambda_L$ rises from $0$ (Gaussian) to
$\approx0.29$ ($\nu=2$), the linear correlation barely moves (it stays near $0.29$), yet the
probability that all three regions fall below their fifth hazard percentile in the same season
roughly quadruples, from about $0.0017$ to $0.0073$. A correlation-only diagnostic would
declare these portfolios equivalent; they are not.

\paragraph{Heavier tail dependence returns the deep tail to the insurer.} Because the joint
catastrophe becomes more frequent, the bounded parametric payouts of the several regions
increasingly trigger \emph{together} in the same severe year, stacking into a large aggregate
payout precisely when independence would have kept the portfolio loss moderate. The right panel
of Figure~\ref{fig:taildep} tracks the $200$-year dominance margin $b^+_{200}-b^-_{200}$ against
$\lambda_L$ at fixed correlation: it declines monotonically for both designs as tail dependence
strengthens---from about $23$ to $13$ for the continuous design and from about $35$ to $17$ for
the binary design over the range studied. The deep tail is being handed back from the insuree
toward the insurer purely through tail dependence, with correlation held constant. This is the
same reabsorption mechanism that the comonotonic limit produced in
Section~\ref{sec:diversification}, but here it is driven by the genuinely relevant quantity and it
operates well before correlation reaches its extreme: even a moderately heavy-tailed coupling
materially shifts who carries the once-in-two-century year.

\paragraph{Implications.} First, the relocation results of this paper should be read against tail
dependence, not correlation: two books with identical correlation and identical AAL-neutral
contracts can allocate the deep tail very differently depending on how their hazards behave in
the joint extreme, and only the tail-dependence view distinguishes them. Second, the binary
design is again the more dependence-sensitive---its margin sits higher but falls faster---because
its fixed payouts stack most mechanically when regions trigger together. Third, the climate
implication sharpens the discussion of Section~\ref{sec:whogains}: climate change is widely
expected to increase not just the marginal severity of extremes but their \emph{spatial
co-occurrence}, i.e.\ $\lambda_L$ itself. By the mechanism here, rising tail dependence pushes
deep-tail cost back toward insurers even as fatter marginal tails push it toward insurees, so the
net incidence of catastrophe risk under NatPar is the resultant of two opposing climate forces,
neither of which is visible to a correlation-based portfolio model. For supervision this reinforces
the case for the occurrence basis $M^B$ and for stress tests specified in terms of joint-extreme
(tail-dependence) scenarios rather than correlation matrices.

\begin{table}[H]
\centering
\caption{Tail dependence versus correlation (continuous and binary, $\beta=0.95$, independent exposures, fixed target correlation $\rho=0.3$). As the lower tail-dependence coefficient $\lambda_L$ increases (via the $t$-copula degrees of freedom $\nu$), the realised correlation is essentially unchanged while the joint cold-extreme probability $\PR(\text{all below }5\%)$ rises and the $200$-year dominance margin $b^+_{200}-b^-_{200}$ falls.}
\label{tab:t10}
\begin{tabular}{lccccrrr}
\toprule
$\nu$ & Design & $\lambda_L$ & $\widehat{\Corr}$ & $\PR(\text{joint }5\%)$ & $b^+_{200}$ & $b^-_{200}$ & margin \\
\midrule
Gauss & c & 0.00 & 0.30 & 0.0017 & 88.80 & 50.04 & 38.76 \\
Gauss & b & 0.00 & 0.30 & 0.0017 & 90.22 & 49.70 & 40.52 \\
8 & c & 0.06 & 0.29 & 0.0030 & 84.80 & 50.03 & 34.77 \\
8 & b & 0.06 & 0.29 & 0.0030 & 88.22 & 52.43 & 35.79 \\
4 & c & 0.16 & 0.29 & 0.0044 & 83.38 & 51.16 & 32.21 \\
4 & b & 0.16 & 0.29 & 0.0044 & 86.91 & 56.88 & 30.03 \\
2 & c & 0.29 & 0.28 & 0.0073 & 82.47 & 55.89 & 26.58 \\
2 & b & 0.29 & 0.28 & 0.0073 & 85.67 & 60.43 & 25.24 \\
\bottomrule
\end{tabular}
\end{table}

\subsection{A testable hypothesis, not yet an empirical claim}
We frame the above as a hypothesis that the NatPar reporting framework is built to express and
test, rather than as an established empirical result. The frost case study is a stationary,
single-season illustration; demonstrating the relocation empirically would require a contract
held fixed while exposure and hazard distributions are made non-stationary over time, so that
the crossover horizon can be tracked as the tail fattens. The apparatus needed for that
test---AAL-neutral calibration to hold the mean fixed, basis return-period curves to expose the
timing, and the crossover horizon as a summary statistic---is exactly what the framework
provides. We regard quantifying the relocation under explicit climate and exposure-growth
scenarios as the natural next application, and note that the occurrence-basis object $M^B$ of
Section~\ref{sec:natpar} is the appropriate lens for supervisors, since it captures the worst
single-region shortfall in a year rather than a portfolio average that may net out.
\section{Extensions and discussion}\label{sec:ext}
The frost example shows, in a controlled setting, how NatPar emerges from
an existing NatCat model and how it changes the distribution of losses and capital-relevant
metrics. We sketch several extensions and discuss broader implications and limitations.

\subsection{Multi-trigger and multi-peril NatPar programmes}
Real-world parametric programmes often involve multiple indices, multiple perils and layered
structures. Within a single peril it is natural to consider multi-trigger structures using several
aspects of the hazard, $P_{e,m} = I_m(Z^{(1)}_e, \dots, Z^{(J)}_e)$ with $J>1$ indices per event. The
counting argument of Section~\ref{sec:pricing} gives a clean design rule here: each additional
contractible index is also an additional free parameter, and so buys one more exactly-satisfiable
moment condition (e.g.\ AAL-neutrality \emph{and} tail-neutrality). Multi-trigger designs are
attractive when they capture qualitatively different damage channels difficult to encode in a
single index, and become problematic when the payout surface is opaque or when small
parameter changes produce large, unintuitive payout changes. If $P$ indexes perils and $R$
indexes regions, a general NatPar portfolio is $S^{\Par} = \sum_{p\in P}\sum_{r\in\R} S^{\Par}_{p,r}$,
driven by the same event-level hazard simulations that drive the NatCat model.

\paragraph{Complexity vs transparency.} NatPar provides a counterweight to the temptation to
exploit the full flexibility of $I(X)$: because the same NatCat models estimate the incremental
reduction in basis risk from each added index, one can compare the benefit (in basis-risk metrics)
against the cost (transparency, operational complexity, regulatory acceptability). In many cases
a small number of well-chosen simple triggers dominates more elaborate designs.

\subsection{NatPar and solvency/climate-risk regulation}
Regulatory capital regimes typically impose risk measures such as VaR or TVaR at high
confidence over one-year horizons. As shown in the frost example, NatPar portfolios often have
shorter tails and lower TVaR for a given AAL than their indemnity counterparts, due to bounded,
index-linked liabilities and the absence of a long development tail; under tail-sensitive capital
regimes they can therefore be more capital-efficient per unit of expected loss. This raises both
opportunities and questions: parametric structures may allow insurers to maintain coverage in
climate-exposed lines while respecting capital requirements, but they introduce basis risk that
must be understood and, in some cases, explicitly reported or capitalised. Climate-risk stress
testing typically alters the hazard module and re-runs the model; in a NatPar setting the same
altered hazards feed directly into index distributions and parametric payouts, and because
NatPar is more tightly linked to hazard and less to exposure it can give sharper insight into how
balance sheets respond to physical climate shifts. The NatPar framework also offers a way to
present parametric programmes in a language familiar to regulators---as variants of existing
NatCat models with clear, quantifiable basis-risk profiles---and suggests that guidance may wish
to distinguish simple hazard-linked structures from highly complex multi-trigger designs.

\subsection{Limitations and practical challenges}
\paragraph{Data and model limitations.} NatPar relies on the same underlying hazard, exposure
and vulnerability models as NatCat. If these are poor, biased or unstable, the resulting basis-risk
analysis and parametric designs will also be unreliable. In some markets hazard data are sparse,
exposure data incomplete, and loss histories short; in others, non-stationarity is so pronounced
that historical patterns may be a poor guide to the future. NatPar does not remove these
problems; it re-allocates them. In particular, the empirical calibration here uses a Gaussian fit
to 65 seasonal observations per region and a stylised lognormal exposure; richer marginals and
dependence (e.g.\ copulas, or $A_r\mid T_r$ regression) can be substituted without changing the
NatCat--NatPar logic.

\paragraph{Governance of indices and triggers.} Parametric payouts are only as good as the
indices they rely on. This raises questions of governance: who controls the index, how revisions or
errors are handled, how disputes are resolved, and what happens when measurement systems
change. Robust contractual and institutional arrangements are needed to keep indices trustworthy
over the lifetime of a programme.

\paragraph{Market acceptance and client understanding.} Even simple parametric structures can be
unfamiliar to policyholders and intermediaries used to indemnity insurance. Basis risk in
particular can be hard to explain: clients may perceive a contract as unfair if it fails to pay in a
year when they experience losses, even if the design is actuarially sound. Education, clear
documentation, and careful trigger choice are essential for market acceptance.

\paragraph{Operational and legal considerations.} Implementing NatPar programmes at scale
requires robust operational processes: index calculation and verification, payout automation,
integration with underwriting and claims systems, and coordination with reinsurance and capital
market transactions. Legal frameworks for parametric products may be less well-developed than
for traditional insurance in some jurisdictions.

\section{Conclusion}\label{sec:summary}
Starting from a simple NatCat model of seasonal frost damage in three simulated regions sharing one
identical damage function, we defined an indemnity cover and constructed
two NatPar alternatives whose payouts depend only on temperature: a continuous design
$P=qD(T)$ and a binary tail-trigger design $P=q\mathbf{1}\{T<\tau\}$, calibrated by AAL-neutrality
and priced by minimising a loss of mismatch. Because the designs share the same hazard and
vulnerability base, differences in EP curves, VaR, and basis behaviour are attributable to contract
structure rather than to inconsistent modelling inputs, and several conclusions emerge that are
generic enough to set a standard for how NatPar programmes should be evaluated.

The first is that payout shape, not the parametric label, governs tail capital. ``Parametric'' is not
automatically ``capital-light'': continuous bounded schedules tend to cap liabilities and dampen
exposure-driven tails, whereas digital schedules can concentrate losses into capital-relevant
quantiles, so trigger probability and payout shape must be judged jointly against the chosen capital
metric. The second is that basis risk is the operational interface between NatCat and NatPar rather
than a nuisance term. Because NatPar replaces the exposure--vulnerability component of indemnity
loss with a low-dimensional hazard-only payout, the residual $B:=Y-P$ is a measurable object, and
the exceedance-basis curves give it a tail view of shortfall and overpayment that is directly suited to
governance and model validation. The third concerns pricing: the trigger price is the solution of
minimising the expected disutility of the two-sided basis, and what moves that price is the curvature
of the penalty, not merely its weight. Asymmetric curvature on shortfall versus overpayment shifts
the premium decisively along the shortfall/overpayment trade-off, while a change of weight under a
linear penalty leaves it at the AAL-neutral value; fair pricing is therefore the symmetric-linear
corner of the problem rather than a competing paradigm, and because the number of conditions a
price can satisfy exactly is bounded by the contract's free parameters, over-determination is cured by
adding a parameter rather than by fighting the algebra.

Beyond the single risk, the portfolio analysis shows that who bears the catastrophe tail is not a fixed
property of a contract but depends on the book it sits in and on how its hazards behave jointly. Under
an AAL-neutral design the mismatch is relocated across time rather than on average: the insuree
tends to gain at short and moderate horizons, where bounded over-payments exceed realised losses,
while the insurer sheds the deep tail past a crossover horizon of order a century. Diversification
across weakly dependent regions pulls that crossover earlier, because the frequent over-payments net
out while the severe shortfalls do not. The direction reverses, however, once the hazards exhibit tail
dependence: when regions reach their cold extremes jointly the bounded payouts stack and the insurer
reabsorbs the deep tail, an effect driven by tail dependence at fixed correlation and therefore
invisible to a correlation-based portfolio model. The net incidence of catastrophe risk under NatPar
is thus the resultant of two opposing climate forces---fatter marginal tails pushing the deep tail
toward the insuree, rising spatial co-occurrence of extremes pushing it back toward the insurer---which
is why we regard the occurrence basis and joint-extreme stress scenarios, rather than averages and
correlation matrices, as the right supervisory lenses.

Taken together, these results support a disciplined reporting template that extends the familiar
NatCat outputs to the parametric portfolio and pairs them with basis-risk exceedance diagnostics and
the loss-minimisation price, making NatPar programmes comparable, auditable, and amenable to
supervision while discouraging gratuitous complexity, since additional triggers and indices should be
justified by measured basis-risk reduction net of the transparency and operational robustness they
cost. Natural Parametric insurance, on this reading, is not a departure from NatCat practice but a
reorganisation of it in response to climate-driven uninsurability, exposure uncertainty, and capital
pressure---and making that reorganisation explicit gives theorists, practitioners, and supervisors a
common language in which to design, analyse, and govern parametric programmes in a way that is both
mathematically coherent and grounded in how catastrophe risk is actually managed.

\appendix
\section{Proof of equation~\eqref{eq:contover}}
Fix $x \ge 0$ and condition on $D$.
\emph{Case 1 (no loss under deductible, $AD \le l$):} then $Y = 0$ and $P - Y = qD'$, so
overpayment exceedance is $qD' > x$, which depends only on hazard.
\emph{Case 2 (above deductible, $AD > l$):} then $Y = AD - l$ and $P - Y = qD' - (AD - l) = l + qD' - AD$,
so $P - Y > x \Leftrightarrow A < (l - x + qD')/D$, together with $A > l/D$. Combining the two cases,
\begin{align*}
\BEP^{c,-}(x) &= \PR(P - Y > x) = \E[\mathbf{1}\{P-Y>x\}\mathbf{1}\{D>0\}] \\
&= \E\!\left[F_A\!\left(\tfrac{l}{D}\right)\mathbf{1}\{D>0,\,qD'>x\}\right] + \E\!\left[\Big(F_A\!\big(\tfrac{l+qD'-x}{D}\big) - F_A\!\big(\tfrac{l}{D}\big)\Big)\mathbf{1}\{D>0,\,qD'>x\}\right] \\
&= \E\!\left[F_A\!\left(\tfrac{l+qD'-x}{D}\right)\mathbf{1}\{qD'>x\}\right]
= \E\!\left[F_A\!\left(\tfrac{l - qd - x}{D}+q\right)\mathbf{1}\{D>x/q+d\}\right].
\end{align*}

\section{Proof of equation~\eqref{eq:binshort}}
\begin{align*}
\BEP^{b,+}(x) &= \PR(Y > x, D \le d) + \PR(Y > x + q, D > d) \\
&= \PR(AD > x + l, D \le d) + \PR(AD > x + q + l, D > d) \\
&= \E\!\left[\overline{F}_A\!\left(\tfrac{x+l}{D}\right)\mathbf{1}\{D\le d\}\right] + \E\!\left[\overline{F}_A\!\left(\tfrac{x+q+l}{D}\right)\mathbf{1}\{D> d\}\right].
\end{align*}

\section{Proof of equation~\eqref{eq:binover}}
\begin{align*}
\BEP^{b,-}(x) &= \PR(-B > x) = \PR(q > x, D > d, Y = 0) + \PR(q - Y > x, D > d, Y > 0) \\
&= \PR\!\left(q > x, D > d, A \le \tfrac{l}{D}\right) + \PR\!\left(q > x, \tfrac{q+l-x}{D} > A > \tfrac{l}{D}, D > d\right) \\
&= \begin{cases} \E\!\left[F_A\!\left(\tfrac{q+l-x}{D}\right)\mathbf{1}\{D>d\}\right], & x < q, \\ 0, & x \ge q. \end{cases}
\end{align*}

\end{document}